# Unusual plastic deformation and damage features in Titanium: experimental tests and constitutive modeling

Benoit Revil-Baudard[1], Oana Cazacu[1*], Philip Flater[1,2], Nitin Chandola[1], J.L. Alves[3]

[1]Department of Mechanical and Aerospace Engineering, University of Florida, REEF,

1350 N. Poquito Rd., Shalimar, FL, USA.

[2]Air Force Research Laboratory, Eglin, FL, USA

[3] MEMS - Microelectromechanical Systems Research Unit,
Department of Mechanical Engineering, University of Minho, Portugal.

## Abstract

In this paper, we present an experimental study on plastic deformation and damage of polycrystalline pure Ti, as well as modeling of the observed behavior. From the mechanical characterization data, it can be concluded that the material displays anisotropy and tension-compression asymmetry. As concerns damage, the X-ray tomography measurements conducted reveal that damage distribution and evolution in this HCP Ti material is markedly different than in a typical FCC material such as copper. Stewart and Cazacu (2011) anisotropic elastic/plastic damage model is used to describe the behavior. All material parameters involved in this model have a clear physical significance, being related to plastic properties, and are determined based on very few simple mechanical tests. It is shown that this model predicts correctly the anisotropy in plastic deformation, and its strong influence on damage distribution and damage accumulation in Ti. Specifically, for a smooth axisymmetric specimen subject to uniaxial tension, damage initiates at the center of the specimen and is diffuse; the level of damage close to failure is very low. On the other hand, for a notched specimen subject to the same loading, the model predicts that damage initiates at the outer surface of the specimen, and further grows from the outer surface to the center of the specimen, which corroborates with the in-situ tomography data.

[*] Corresponding author: Tel: +1(850) 833-9350; fax +1(850) 833-9366; E-mail:cazacu @reef.ufl.edu (Oana Cazacu).

# 1. Introduction

Hexagonal close-packed (HCP) dominated titanium materials are known to display at room temperature plastic anisotropy and a strong tension–compression asymmetry (e.g. for high-purity HCP−Ti, see data reported by Nixon et al. (2010a;b); for TiAl6V, see data of Gilles et al. (2011)). This tension-compression asymmetry at macroscopic level is a consequence of the operation of specific crystallographic deformation mechanisms. Concerning the identification of crystallographic deformation mechanisms operational in high-purity Ti, the reader is referred to the studies of Salem et al. (2003), Knezevic et al. (2013) for monotonic uniaxial loadings, Bouvier et al. (2012) for simple shear monotonic, and cyclic loadings; concerning commercial-purity Ti, see for e.g. Chun et al. (2005), Benhenni et al. (2013), etc.

As concerns modeling the mechanical behavior at macroscopic level, it is generally agreed that classical plasticity models such as von Mises or Hill (1948) cannot capture the specificities of the plastic deformation in Ti materials (e.g. Kuwabara et al (2001), Nixon et al. (2010a)). As pointed out in the review of Banerjee and Williams (2013), recently developed macroscopic level elastic/plastic constitutive models (e.g. Cazacu and Barlat, 2004, Cazacu et al. 2006, Nixon et al. 2010a) are able to capture the main features of the plastic deformation of HCP−titanium. Specifically, the capabilities of these models to predict both the anisotropy and tension-compression asymmetry in plastic deformation have been demonstrated for both low-strain rates loadings (e.g. for pure Ti under uniaxial tension, uniaxial compression, bending, see Nixon et al., (2010b); under torsion, see Revil-Baudard et al., (2014); for TiAl6V under tension, compression, and plane-strain tension, see Gilles et al. (2013) ), and high-rate and impact loadings (e.g. Revil-Baudard et al., 2015).

At present, data concerning damage in titanium materials are extremely limited and alloy dependent. Although titanium materials are anisotropic, the data reported concern only one loading orientation (e.g. Huez et al., 1998).

Moreover, the description of the room-temperature damage and failure of titanium materials is done using either empirical laws for the strain at failure, or evolution laws for the rate of void growth (e.g. Rice and Tracey (1969) law and its various modifications). However, Rice and Tracey (1969) void growth law or the most widely used criteria for ductile damage (e.g. Gurson (1977) and its various modifications) cannot realistically describe damage in Ti materials because the core hypothesis of such models is that the plastic behavior of the matrix is governed by the isotropic von Mises criterion, which is a criterion critically inadequate for Ti materials.

Using rigorous upscaling techniques based on Hill-Mandel lemma (Hill, 1967; Mandel, 1972), Stewart and Cazacu (2011) recently developed an analytic anisotropic model for porous polycrystals that accounts for the combined effects of tension-compression asymmetry and evolving anisotropy on damage. The unusual properties of the yield loci of porous HCP predicted by this model were validated by full-field crystal plasticity simulations (see Lebensohn and Cazacu, 2012).

In this paper, we present an experimental study on plastic deformation and damage of polycrystalline pure Ti, as well as modeling of the observed behavior. The Stewart and Cazacu (2011) model, which will be used for prediction of plastic deformation and damage evolution in this Ti material is presented in Section 2. All material parameters have a clear physical significance, being related to the plastic properties of the material, and their identification is done based on a few characterization tests in uniaxial tension on flat specimens and uniaxial compression on cylindrical specimens (see Sections 3.1-3.2).

To further study damage in the Ti material, additional uniaxial tension tests on axisymmetric cylindrical specimens of circular cross-section were conducted. Both smooth and notched geometries were considered. X-ray micro-computed tomography (XCMT) measurements both ex-situ and in-situ were conducted. Comparison between data and FE predictions of both local and global deformation, and porosity evolution obtained with the model are presented (Section 4). We conclude with a summary of the main experimental findings, and conclusions concerning the capabilities of the model to capture the key features of plasticity-damage couplings in Ti (see Section 5).

2. **Elastic-plastic damage model**

The total rate of deformation $\mathbf{D}$ (the symmetric part of $\dot{\mathbf{F}}\mathbf{F}^{-1}$ where $\mathbf{F}$ is the deformation gradient) is considered to be the sum of an elastic part and a plastic part $\mathbf{D}^{\mathrm{p}}$. The elastic response is described as

$$\dot{\boldsymbol{\sigma}} = \mathbf{C}^{\mathrm{e}} : (\mathbf{D} - \mathbf{D}^{\mathrm{p}}) \qquad (1)$$

where $\dot{\boldsymbol{\sigma}}$ is the Green-Naghdi derivative (see, Green and Naghdi, 1965, ABAQUS, 2009) of the Cauchy stress tensor $\boldsymbol{\sigma}$, $\mathbf{C}^{\mathrm{e}}$ is the fourth-order stiffness tensor while ":" denotes the doubled contracted product between the two tensors.

The plastic strain rate is defined with respect to the stress potential $\varphi$ as:

$$\mathbf{D}^{\mathrm{p}} = \dot{\lambda}\frac{\partial \varphi}{\partial \boldsymbol{\sigma}} \qquad (2)$$

where $\dot{\lambda}$ is the plastic multiplier.

To describe the behavior of a porous anisotropic solid with matrix being incompressible but displaying tension-compression asymmetry, we will Stewart and Cazacu (2011) potential. This plastic potential (which coincides with the yield function of the porous solid) was derived using the kinematic non-linear homogenization approach of Hill (1967) and Mandel (1972), assuming that the matrix material is governed by the quadratic form of the orthotropic yield criterion of Cazacu et al.(2006). It has the following expression:

$$\varphi(\boldsymbol{\sigma}, f) = \hat{m}^2 \frac{\sum_{i=1}^{3}(|\hat{\sigma}_i| - k\hat{\sigma}_i)^2}{\bar{\sigma}_x^T} + \bar{\sigma}_x^T \left( 2f \cosh\left(\frac{3\sigma_m}{h\bar{\sigma}_x^T}\right) - (1+f^2) \right) = 0, \tag{3}$$

where k is a parameter describing the tension-compression asymmetry of the matrix; f is the void volume fraction, and $\hat{\sigma}_1, \hat{\sigma}_2, \hat{\sigma}_3$ are the principal values of the transformed stress tensor

$$\hat{\boldsymbol{\sigma}} = \mathbf{L} : \boldsymbol{\sigma}'. \tag{4}$$

In Eq. (3) $\boldsymbol{\sigma}'$ is the deviator of the Cauchy stress tensor $\boldsymbol{\sigma}$, ($\boldsymbol{\sigma}' = \boldsymbol{\sigma} - \sigma_m \mathbf{I}$, $\sigma_m = (1/3)\boldsymbol{\sigma} : \mathbf{I}$, with $\mathbf{I}$ being the second-order identity tensor), $\mathbf{L}$ is a fourth-order symmetric tensor describing the anisotropy of the matrix, while ":" denotes the doubled contracted product between the two tensors.

Let (**x,y,z**) be the reference frame associated with orthotropy. In the case of a plate, **x**, **y** and **z** represent the rolling, transverse and normal directions, respectively. Relative to the orthotropy axes, the fourth-order tensor **L** is represented in Voigt notation by:

$$\mathbf{L} = \begin{bmatrix} L_{11} & L_{12} & L_{13} & & & \\ L_{12} & L_{22} & L_{23} & & & \\ L_{13} & L_{23} & L_{33} & & & \\ & & & L_{44} & & \\ & & & & L_{55} & \\ & & & & & L_{66} \end{bmatrix} \tag{5}$$

In the expression of $\varphi(\sigma, f)$ given by Eq. (3), $\sigma_x^T$ is the uniaxial tensile yield stress along an axis of orthotropy of the fully-dense material (i.e., the direction **x**) while $\hat{m}$ is a constant which depends on the anisotropy coefficients $L_{ij}$ and the strength differential parameter k, i.e.:

$$\hat{m} = \left[ \left(|\Phi_1| - k\Phi_1\right)^2 + \left(|\Phi_2| - k\Phi_2\right)^2 + \left(|\Phi_3| - k\Phi_3\right)^2 \right]^{-1/2}, \tag{6}$$

where

$$\Phi_1 = (2L_{11} - L_{12} - L_{13})/3, \ \Phi_2 = (2L_{12} - L_{22} - L_{23})/3, \ \Phi_3 = (2L_{13} - L_{23} - L_{33})/3. \tag{7}$$

The parameter h in the potential $\varphi(\sigma, f)$ depends on the matrix anisotropy and the sign of the mean stress, $\sigma_m$. Its expression is:

$$h = \sqrt{n(4t_1 + 6t_2)/5}, \tag{8}$$

with

$$n = \begin{cases} \dfrac{3}{\hat{m}^2 \left(3k^2 - 2k + 3\right)} & \text{if } \sigma_m < 0, \\[2ex] \dfrac{3}{\hat{m}^2 \left(3k^2 + 2k + 3\right)} & \text{if } \sigma_m \geq 0. \end{cases} \tag{9}$$

The scalars $t_1$ and $t_2$ involved in the expression of h account for the anisotropy of the fully-dense material, and are expressed as:

$$t_1 = 3\left(B_{13}B_{23} + B_{12}B_{23} + B_{12}B_{13} + 2B_{12}^2 + 2B_{13}^2 + 2B_{23}^2\right) \tag{10a}$$

$$t_2 = B_{44}^2 + B_{55}^2 + B_{66}^2 \tag{10b}$$

In the above equation, $B_{ij}$, with $i,j = 1...3$, are the components of the inverse of the tensor $\mathbf{C} = \mathbf{LK}$, $\mathbf{K}$ denoting the 4$^{\text{th}}$ order deviatoric unit tensor

$$\mathbf{K} = \begin{bmatrix} 2/3 & -1/3 & -1/3 & & & \\ -1/3 & 2/3 & 1/3 & & & \\ -1/3 & -1/3 & 2/3 & & & \\ & & & 1 & & \\ & & & & 1 & \\ & & & & & 1 \end{bmatrix} \quad (11)$$

The expressions of the components of $\mathbf{B}$ in terms of the matrix anisotropy coefficients (i.e. components of the orthotropy tensor $\mathbf{L}$) are given in Appendix A.

In the expression of $\varphi(\boldsymbol{\sigma}, f)$ given by Eq. (3), $\bar{\sigma}_x^T$ stands for the tensile effective stress of the fully-dense material in one of the directions of orthotropy of the material, generally in the rolling direction $\mathbf{x}$. Hardening of the matrix is considered to be governed by the effective plastic strain, $\bar{\varepsilon}^p$. The rate of the effective plastic strain $\dot{\bar{\varepsilon}}^p$ is obtained, assuming the equivalence of microscopic and macroscopic inelastic work and associated flow rule, as

$$\bar{\sigma} \dot{\bar{\varepsilon}}^p (1-f) = \boldsymbol{\sigma} : \mathbf{D}^p, \quad (12)$$

so,

$$\dot{\bar{\varepsilon}}^p = \frac{\boldsymbol{\sigma} : \mathbf{D}^p}{(1-f)\bar{\sigma}} = \dot{\lambda} \frac{\boldsymbol{\sigma} : \dfrac{\partial \varphi}{\partial \boldsymbol{\sigma}}}{(1-f)\bar{\sigma}}. \quad (13)$$

The rate of change of the void volume fraction ($\dot{f}$) is considered to result from the growth of existing voids and the nucleation of new ones. Void growth is obtained from mass conservation and the use of the plastic flow rule (Eq.(2)) in conjunction with Eq. (3). Void nucleation is considered to be due to plastic strain, as suggested by Gurson (1975) based on Gurland's (1972) experimental data, and to the mean stress, $\sigma_m$, as discussed in Argon et al. (1975). Both plastic strain controlled nucleation and mean stress controlled nucleation are considered to follow a normal distribution with mean value ($\varepsilon_N$, $\sigma_P$) and a standard deviation ($s_N$, $s_P$), as proposed by Chu and Needleman (1980). Thus,

$$\dot{f} = (1-f)\mathbf{D}^p:\mathbf{I} + A_N \dot{\bar{\varepsilon}}^p + B_N \dot{\sigma}_m, \tag{14}$$

where

$$A_N = \frac{f_N}{s_N \sqrt{2\pi}} \exp\left[-\frac{1}{2}\left(\frac{\bar{\varepsilon}^p - \varepsilon_N}{s_N}\right)^2\right],$$

$$B_N = \begin{cases} 0 & \text{if } \dot{\sigma}_m < 0 \\ \dfrac{f_P}{s_P \sqrt{2\pi}} \exp\left[-\dfrac{1}{2}\left(\dfrac{\sigma_m - \sigma_P}{s_P}\right)^2\right] & \text{if } \dot{\sigma}_m > 0 \end{cases}.$$

In the above equation, $f_N$ and $f_P$ represent the volume fraction of voids nucleated by the plastic strain, and by the mean stress, respectively.

It is to be noted that in the absence of voids ($f = 0$), Stewart and Cazacu (2011) criterion $\varphi(\sigma, f)$, given by Eq. (3), reduces to that of the matrix, i.e. the quadratic form of Cazacu et al. (2006) orthotropic criterion:

$$\tilde{\sigma}_e = \sigma_x^T, \tag{15}$$

with

$$\tilde{\sigma}_e = \hat{m}\sqrt{\sum_{i=1}^{3}\left(\left|\hat{\sigma}_i\right| - k\hat{\sigma}_i\right)^2}.$$

Note that the Cazacu et al. (2006) effective stress, $\tilde{\sigma}_e$, accounts for the tension-compression asymmetry of the matrix through the parameter k, and depends on all the invariants of the stress deviator, $\boldsymbol{\sigma}'$, as well as on the mixed invariants of $\boldsymbol{\sigma}'$ and the symmetry tensors associated with orthotropy, namely $\mathbf{M}_1 = \mathbf{x} \otimes \mathbf{x}$, $\mathbf{M}_2 = \mathbf{y} \otimes \mathbf{y}$, $\mathbf{M}_3 = \mathbf{z} \otimes \mathbf{z}$ (see Cazacu et al., 2006). Thus, the anisotropic yield function $\varphi(\boldsymbol{\sigma}, f)$ for the porous material given by Eq. (3) accounts for the combined effects of the matrix orthotropy and tension-compression asymmetry. Furthermore, since the hydrostatic parameter h depends on the anisotropy coefficients $L_{ij}$, on the strength-differential parameter k, and on the sign of the applied mean stress $\sigma_m$ (see Eq. (8)-(10)), Stewart and Cazacu (2011) criterion (3) accounts for the combined effects of the tension-compression asymmetry and orthotropy on the dilatational response. In particular, according to this criterion, the yield surface of the porous solid does not display the centro-symmetry properties of the yield surface of a porous von Mises solid (see Cazacu et al. (2013)) or of a porous Tresca solid (see Cazacu et al. (2014)). Specifically, the yield limit for purely tensile hydrostatic loading is different than the yield limit for purely compressive hydrostatic loading. Indeed, according to the criterion (3), for tensile hydrostatic loading, yielding of the porous material occurs when $\sigma_m = \hat{p}_Y^+$, with

$$\hat{p}_Y^+ = -\frac{\sigma_x^T}{3}\sqrt{\frac{3}{\hat{m}^2(3k^2 + 2k + 3)}}\left(\frac{4t_1 + 6t_2}{5}\right)\ln(f), \qquad (16a)$$

whereas for compressive hydrostatic loading, yielding occurs when $\sigma_m = \hat{p}_Y^-$, with

$$\hat{p}_Y^- = \frac{\sigma_x^T}{3} \sqrt{\frac{3}{\hat{m}^2(3k^2 - 2k + 3)}\left(\frac{4t_1 + 6t_2}{5}\right)} \ln(f), \quad (16b)$$

the expressions for $\hat{m}$, $t_1$, and $t_2$ in terms of the orthotropy coefficients and k being given by Eqs. (6), (10a)-(10b), respectively. Furthermore, according to the Stewart and Cazacu (2011) criterion, the yield function of the porous solid is no longer invariant with respect to the transformation $(\sigma_m, \boldsymbol{\sigma}') \to (\sigma_m, -\boldsymbol{\sigma}')$. To further illustrate that $\varphi(\boldsymbol{\sigma}, f)$ depends on the third-invariant of the stress deviator, $J_3 = \sigma'_1 \sigma'_2 \sigma'_3$, and on the mixed invariants associated with orthotropy, in the following we present its expressions corresponding to axisymmetric loading about the **x**-axis, **y**-axis and **z**-axis of orthotropy, respectively. In each case, $\sigma_1$ will denote the axial stress while $\sigma_3$ will denote the lateral pressure (the value of the principal stresses that are equal). Thus, irrespective of the orientation of the reference system associated with loading with respect to the material axes (**x**, **y**, **z**), the von Mises equivalent stress is $\sigma_e = |\sigma_1 - \sigma_3|$, the mean stress $\sigma_m = (\sigma_1 + 2\sigma_3)/3$, and the deviatoric third invariant is $J_3 = 2(\sigma_1 - \sigma_3)^3/27$.

Let first consider the case of axisymmetric loading about **x**-axis i.e., $\boldsymbol{\sigma} = \sigma_1(\mathbf{x} \otimes \mathbf{x}) + \sigma_3(\mathbf{y} \otimes \mathbf{y} + \mathbf{z} \otimes \mathbf{z})$.

- For $J_3 \leq 0$ (i.e. minor principal stress along **x**-axis) Eq. (3) becomes:

$$\varphi = \begin{cases} \left(\frac{\sigma_x^T}{\sigma_x^C}\right)^2 \left(\frac{\sigma_e}{\sigma_x^T}\right)^2 + 2f \cosh\left(\frac{\sigma_m}{\sigma_x^T} \sqrt{\frac{15\hat{m}^2(3k^2 - 2k + 3)}{4t_1 + 6t_2}}\right) - (1 + f^2), & \text{if } \sigma_m < 0, \\ \left(\frac{\sigma_x^T}{\sigma_x^C}\right)^2 \left(\frac{\sigma_e}{\sigma_x^T}\right)^2 + 2f \cosh\left(\frac{\sigma_m}{\sigma_x^T} \sqrt{\frac{15\hat{m}^2(3k^2 + 2k + 3)}{4t_1 + 6t_2}}\right) - (1 + f^2), & \text{if } \sigma_m \geq 0 \end{cases} \quad (17a)$$

while for $J_3 \geq 0$ (i.e. major principal stress along the **x**-axis), the expression of the criterion (3) is:

$$\varphi = \begin{cases} \left(\dfrac{\sigma_e}{\sigma_x^T}\right)^2 + 2f \cosh\left(\dfrac{\sigma_m}{\sigma_x^T}\sqrt{\dfrac{15\hat{m}^2(3k^2 - 2k + 3)}{4t_1 + 6t_2}}\right) - (1+f^2), & \text{if } \sigma_m < 0, \\ \left(\dfrac{\sigma_e}{\sigma_x^T}\right)^2 + 2f \cosh\left(\dfrac{\sigma_m}{\sigma_x^T}\sqrt{\dfrac{15\hat{m}^2(3k^2 + 2k + 3)}{4t_1 + 6t_2}}\right) - (1+f^2), & \text{if } \sigma_m \geq 0. \end{cases} \quad (17b)$$

with $\sigma_x^T$ the matrix uniaxial tensile yield stress and $\sigma_x^C$ the matrix uniaxial compressive yield stress along the **x**-axis.

For axisymmetric loading about **y**-axis: $\boldsymbol{\sigma} = \sigma_1(\mathbf{y} \otimes \mathbf{y}) + \sigma_3(\mathbf{x} \otimes \mathbf{x} + \mathbf{z} \otimes \mathbf{z})$, Stewart and Cazacu (2011) criterion writes:

- for $J_3 \leq 0$ (i.e. minor principal stress along **y**-axis):

$$\varphi = \begin{cases} \left(\dfrac{\sigma_x^T}{\sigma_y^C}\right)^2\left(\dfrac{\sigma_e}{\sigma_x^T}\right)^2 + 2f \cosh\left(\dfrac{\sigma_m}{\sigma_x^T}\sqrt{\dfrac{15\hat{m}^2(3k^2 - 2k + 3)}{4t_1 + 6t_2}}\right) - (1+f^2), & \text{if } \sigma_m < 0, \\ \left(\dfrac{\sigma_x^T}{\sigma_y^C}\right)^2\left(\dfrac{\sigma_e}{\sigma_x^T}\right)^2 + 2f \cosh\left(\dfrac{\sigma_m}{\sigma_x^T}\sqrt{\dfrac{15\hat{m}^2(3k^2 + 2k + 3)}{4t_1 + 6t_2}}\right) - (1+f^2), & \text{if } \sigma_m \geq 0. \end{cases} \quad (18a)$$

- for $J_3 \geq 0$ (i.e. major principal stress along the **y**-axis), the expression of the criterion (3) is

$$\varphi = \begin{cases} \left(\dfrac{\sigma_x^T}{\sigma_y^T}\right)^2\left(\dfrac{\sigma_e}{\sigma_x^T}\right)^2 + 2f\cosh\left(\dfrac{\sigma_m}{\sigma_x^T}\sqrt{\dfrac{15\hat{m}^2(3k^2 - 2k + 3)}{4t_1 + 6t_2}}\right) - (1+f^2), & \text{if } \sigma_m < 0, \\ \left(\dfrac{\sigma_x^T}{\sigma_y^T}\right)^2\left(\dfrac{\sigma_e}{\sigma_T}\right)^2 + 2f\cosh\left(\dfrac{\sigma_m}{\sigma_x^T}\sqrt{\dfrac{15\hat{m}^2(3k^2 + 2k + 3)}{4t_1 + 6t_2}}\right) - (1+f^2), & \text{if } \sigma_m \geq 0. \end{cases} \quad (18b)$$

where $\sigma_y^C$ and $\sigma_y^T$ denote the matrix uniaxial compressive and tensile yield stresses along the **y** axis, respectively.

For the case of axisymmetric loading along the **z**-axis, i.e. $\boldsymbol{\sigma} = \sigma_1(\mathbf{z}\otimes\mathbf{z}) + \sigma_3(\mathbf{x}\otimes\mathbf{x} + \mathbf{y}\otimes\mathbf{y})$, Stewart and Cazacu (2011) criterion writes:

- for $J_3 \leq 0$ (i.e. minor principal stress along the **z**-axis):

$$\varphi = \begin{cases} \left(\dfrac{\sigma_x^T}{\sigma_z^C}\right)^2\left(\dfrac{\sigma_e}{\sigma_x^T}\right)^2 + 2f\cosh\left(\dfrac{\sigma_m}{\sigma_x^T}\sqrt{\dfrac{15\hat{m}^2(3k^2 - 2k + 3)}{4t_1 + 6t_2}}\right) - (1+f^2), & \text{if } \sigma_m < 0, \\ \left(\dfrac{\sigma_x^T}{\sigma_z^C}\right)^2\left(\dfrac{\sigma_e}{\sigma_T}\right)^2 + 2f\cosh\left(\dfrac{\sigma_m}{\sigma_x^T}\sqrt{\dfrac{15\hat{m}^2(3k^2 + 2k + 3)}{4t_1 + 6t_2}}\right) - (1+f^2), & \text{if } \sigma_m \geq 0. \end{cases} \quad (19a)$$

while

- for $J_3 \geq 0$ (i.e. major principal stress along the **z**-axis), the expression of the criterion (3) is

$$\varphi = \begin{cases} \left(\dfrac{\sigma_x^T}{\sigma_z^T}\right)^2\left(\dfrac{\sigma_e}{\sigma_x^T}\right)^2 + 2f\cosh\left(\dfrac{\sigma_m}{\sigma_x^T}\sqrt{3\hat{m}^2(3k^2 - 2k + 3)\left(\dfrac{5}{4t_1 + 6t_2}\right)}\right) - (1+f^2), & \text{if } \sigma_m < 0, \\ \left(\dfrac{\sigma_x^T}{\sigma_z^T}\right)^2\left(\dfrac{\sigma_e}{\sigma_x^T}\right)^2 + 2f\cosh\left(\dfrac{\sigma_m}{\sigma_x^T}\sqrt{3\hat{m}^2(3k^2 + 2k + 3)\left(\dfrac{5}{4t_1 + 6t_2}\right)}\right) - (1+f^2), & \text{if } \sigma_m \geq 0. \end{cases} \quad (19b)$$

where $\sigma_z^C$ and $\sigma_z^T$ denote the matrix uniaxial compressive and tensile yield stresses along the **z** axis (i.e. the through thickness direction), respectively.

In summary, Stewart and Cazacu (2011) criterion has different expressions depending on the orientation of the principal directions of stress with respect to the material axes, the ordering of the principal stresses and the sign of the mean stress. The combined effects of the matrix tension-compression asymmetry and its orthotropy are evident by comparing expressions (17)-(19) of the criterion. Note that in the case of axisymmetric loading about the **x**-axis, the sensitivity to the third-invariant $J_3$ results from the matrix tension-compression asymmetry along the **x**-axis (i.e. it's due to $\sigma_x^C \neq \sigma_x^T$), for axisymmetric loading about the **y**-axis, the sensitivity to $J_3$ is due to the matrix tension-compression asymmetry along the **y**-direction of orthotropy (i.e. it's due to $\sigma_y^C \neq \sigma_y^T$), while for axisymmetric loading about the **z**-axis, the sensitivity to $J_3$ results from the matrix tension-compression asymmetry along the **z**-axis (i.e. it's due to $\sigma_z^C \neq \sigma_z^T$). In other words, the sensitivity to $J_3$ is due to the tension-compression asymmetry ratio in the direction of the applied axial stress.

If the matrix is isotropic, i.e. the anisotropy tensor $\mathbf{L} = \mathbf{I_4}$, where $\mathbf{I_4}$ is the 4$^{th}$ order identity tensor ( i.e. $L_{ijkl} = \frac{1}{2}(\delta_{ik}\delta_{jl} + \delta_{il}\delta_{jk})$, with i, j, k, l = 1...3.), then $\Phi_1 = 2/3$, $\Phi_2 = \Phi_3 = -1/3$, (see also Eq. (7)) and the anisotropic Stewart and Cazacu (2011) criterion (Eq. (3)) reduces to Cazacu and Stewart (2009) criterion for isotropic porous solids:

$$m^2 \sum_{i=1}^{3} \left( \frac{|\sigma_i'| - k\sigma_i'}{\sigma_T} \right)^2 + 2f \, \cosh\left( z_s \frac{3\sigma_m}{2\sigma_T} \right) - \left(1 + f^2\right) = 0, \tag{20}$$

the material constant m and $z_s$ depending only on the tension-compression asymmetry of the matrix, i.e.

$$m = \sqrt{\frac{9}{2(3k^2-2k+3)}}$$

$$z_s = \begin{cases} 1 & \text{if } \sigma_m < 0, \\ \sqrt{\dfrac{3k^2 + 2k + 3}{3k^2 - 2k + 3}} & \text{if } \sigma_m \geq 0. \end{cases} \quad (21)$$

## 3. Plastic Deformation in Titanium: characterization tests and model identification

### 3.1. Experimental results in uniaxial compression and uniaxial tension

The material used in this work was high-purity (99.9%) titanium, and it was purchased from Tico Titanium, Inc. The material was supplied in the form of a plate with dimensions 610 x 610 x 16 mm$^3$. Optical microscopy showed that the material has equiaxed grains with an average grain size of about 40 μm. The initial texture of the as received material was measured using Electron Back Scattered Diffraction (EBSD). Analysis of the basal {0001} pole figure shows that the material has a basal texture, with the majority of the grains having their c-axis at 30$^0$ to the normal to the plane of the plate (see Fig. 1).

In order to quantify the influence of the loading direction, and thereby texture on the mechanical response at room-temperature quasi-static tests (nominal strain rate of 0.01 s$^{-1}$) in uniaxial compression were conducted on cylindrical specimens 7.62 mm in diameter by 7.62 mm long that were machined such that the axes of the cylinders were along the rolling direction (RD) and two other in-plane directions, at $45^0$ (DD) and $90^0$ (TD) with respect to RD, respectively. In addition, tests were conducted on specimens with the axis along the normal direction (ND) of the plate. It is to be noted that the stress-strain curves along the in-plane directions are very close, but there is a marked difference between the yield stress in the normal direction and the average in-plane yield stresses (see Fig. 2).

In-plane quasi-static tension tests (nominal strain rate of 0.01 $s^{-1}$) were also conducted on flat dog-bone specimens, the rectangular cross section within the gauge length being 3.2 mm by 1.6 mm. Digital image correlation (DIC) techniques were used to determine the strain fields (axial and width strains) in the gage zone. To assess the repeatability and consistency of the test results, five tests were performed for each orientation. The average true stress-true strain curves in RD, 45°, and TD are shown in Fig. 3. The stress-strain curves are very close. It is worth noting that this titanium material displays a strong and evolving tension-compression asymmetry, irrespective of the in-plane loading orientation, the strength in unixial compression being higher than the strength in unixial tension (see Fig. 4 showing the tension-compression asymmetry of the material for loadings along TD).

Lankford coefficients, also known as r-values, are a measure of the formability of a metallic material. The plastic strain ratio, $r(\theta)$, is defined as the ratio of the width to thickness plastic strain increments measured in a uniaxial tension test along an in-plane direction at an angle $\theta$ with respect to RD, i.e., $r_\theta = \dot{\varepsilon}_{width} / \dot{\varepsilon}_{thickness}$. Generally, the r-values are obtained from measurements of the axial and width strains, the thickness strain being determined assuming plastic incompressibility (for example, see Barlat and Richmond, 1987).

For each specimen orientation, r-values were determined for several levels of plastic strains and are listed in Table 1.

| Orientation θ (deg) | $\bar{\varepsilon}^p=0$ | $\bar{\varepsilon}^p=0.05$ | $\bar{\varepsilon}^p=0.1$ | $\bar{\varepsilon}^p=0.15$ | $\bar{\varepsilon}^p=0.2$ | $\bar{\varepsilon}^p=0.25$ | $\bar{\varepsilon}^p=0.3$ | $\bar{\varepsilon}^p=0.35$ |
|---|---|---|---|---|---|---|---|---|
| 0° | 2.87 | 2.68 | 2.52 | 2.38 | 2.26 | 2.15 | 2.07 | 2.01 |
| 45° | 2.08 | 2.01 | 1.94 | 1.88 | 1.82 | 1.78 | 1.74 | 1.71 |
| 90° | 1.57 | 1.56 | 1.55 | 1.53 | 1.52 | 1.50 | 1.48 | 1.46 |

**Table 1**: Lankford coefficients (r-values) for several in-plane orientations θ measured with respect to the rolling direction. For each orientation, the values corresponding to several levels of accumulated plastic strain are reported.

It is worth noting that at 0.2% offset, irrespective of the orientation, the r-value is very large. Irrespective of the level of strain, the largest r-value is along RD. For comparison, for an isotropic material, r(θ) =1, irrespective of the tensile loading direction θ, for a typical aluminum alloy, e.g. AA 3103-O, the r(θ) range between 0.5 and 0.6 (see Barlat et al., 2004). It should also be noted that for the Ti material studied, irrespective of the orientation, the Lankford coefficients tend to decrease with the accumulated plastic strain (see Table 1).

## 3.2. Identification of the material parameters of Stewart and Cazacu (2011) model

As already mentioned, all material parameters involved in the expression of the criterion $\varphi(\boldsymbol{\sigma}, f)$, namely the anisotropy coefficients $L_{ij}$ as well as the parameter *k* (see Eq. (5)-(11)) are associated to the plastic properties of the material. For the identification of these parameters, the experimental data obtained in unixial tension and compression (yield stresses and r-values) were used. More details on the identification procedure can be found in Cazacu et al. (2006).

To model the difference in hardening rates between tension and compression loadings observed experimentally (e.g., see Fig. 4), all these material parameters are considered to evolve with accumulated plastic deformation. The numerical values of these parameters corresponding to four individual levels of equivalent plastic strains (up to 0.35 strain) are listed in Table 2, the values corresponding to any given level of plastic strain $\bar{\varepsilon}_p^j \leq \bar{\varepsilon} \leq \bar{\varepsilon}_p^{j+1}$ are obtained by linear interpolation, i.e.:

$$
\begin{aligned}
L_{ij}(\bar{\varepsilon}) &= \alpha(\bar{\varepsilon}) L_{ij}(\bar{\varepsilon}_p^j) + (1-\alpha(\bar{\varepsilon})) L_{ij}(\bar{\varepsilon}_p^{j+1}) \\
k(\bar{\varepsilon}) &= \alpha(\bar{\varepsilon}) k(\bar{\varepsilon}_p^j) + (1-\alpha(\bar{\varepsilon})) k(\bar{\varepsilon}_p^{j+1})
\end{aligned}
\tag{22}
$$

The interpolation parameter $\alpha$ involved in Eq. (22) is defined as

$$
\alpha = \frac{\bar{\varepsilon} - \bar{\varepsilon}_p^j}{\bar{\varepsilon}_p^{j+1} - \bar{\varepsilon}_p^j} .
\tag{23}
$$

| Plastic strain | $L_{22}$ | $L_{33}$ | $L_{12}$ | $L_{13}$ | $L_{23}$ | $L_{44}$ | k |
|---|---|---|---|---|---|---|---|
| 0.02 | 0.971 | 1.316 | 0.022 | 0.189 | 0.152 | 0.972 | -0.304 |
| 0.05 | 0.989 | 1.243 | 0.089 | 0.193 | 0.173 | 0.909 | -0.313 |
| 0.1 | 0.992 | 1.046 | 0.016 | 0.075 | 0.053 | 0.983 | -0.363 |
| 0.15 | 0.996 | 0.915 | -0.015 | 0.021 | 0.000 | 1.016 | -0.419 |
| 0.2 | 0.998 | 0.849 | -0.048 | -0.012 | -0.034 | 1.050 | -0.472 |
| 0.25 | 0.998 | 0.815 | -0.089 | -0.041 | -0.068 | 1.092 | -0.518 |
| 0.3 | 0.998 | 0.797 | -0.130 | -0.068 | -0.099 | 1.134 | -0.554 |
| 0.35 | 1.000 | 0.772 | -0.178 | -0.097 | -0.135 | 1.183 | -0.635 |

**Table 2**: Model parameters for the high-purity orthotropic HCP-titanium corresponding to different values of plastic strain, $\bar{\varepsilon}^p$; for any strain level $L_{11}$ is set to unity.

It is considered that the material's hardening is isotropic and it is governed by the equivalent plastic strain according to a power-law:

$$Y(\bar{\varepsilon}^p) = \beta(\varepsilon_0 + \bar{\varepsilon}^p)^n, \qquad (24)$$

In Eq. (24), $\beta$ and $\varepsilon_0$ are material parameters, which can be estimated based on the experimental uniaxial tension axial stress vs. true strain curve in RD. The values of these parameters for the material studied are: $\beta$=413 MPa, $\varepsilon_0$ =0.6445 and n =1.

As shown, the Stewart and Cazacu (2011) criterion for porous solids depends not only on all stress invariants, but also on the mixed invariants associated with orthotropy (see Eq. (17)-(19)). As a consequence, even for the simplest axisymmetric loadings, the yield locus does not display the usual properties of existing criteria for porous solids, whether phenomenological or micro-mechanics-based. To illustrate the combined effects of the anisotropy and tension-compression asymmetry of the plastic flow on yielding of the porous Ti material studied, in Fig. 5 are shown the projections of the yield surface corresponding to an equivalent plastic strain $\bar{\varepsilon}^p$=0.25 and a porosity $f = 0.01$ for axisymmetric loadings where the axial stress is oriented along either the RD (**x**-axis), TD (**y**-axis) or normal direction (ND) (**z**- axis). It is worth noting the strong effect of the third-invariant, $J_3$, on the response of the porous material. Furthermore, this strong influence of the ordering of the principal stresses has a clear explanation, being linked to the asymmetry in plastic deformation. For example, in the case of axisymmetric loading with axial stress along RD-axis and equal lateral stresses $\Sigma_{TD}=\Sigma_{ND}$, the response depends on whether the axial stress $\Sigma_{RD} > \Sigma_{TD}=\Sigma_{ND}$ (i.e. for loading which corresponds to $J_3 > 0$) or $\Sigma_{RD} < \Sigma_{TD}=\Sigma_{ND}$ (i.e. for loading which corresponds to $J_3 < 0$) (see also Eqs. (17)). In particular, for purely deviatoric loadings, the model predicts that the difference in response is exactly the ratio between the RD flow stresses in tension and compression, i.e. $\sigma_x^T/\sigma_x^C$. Since the material studied has $\sigma_x^T/\sigma_x^C < 1$, its overall response is softer for axisymmetric loadings corresponding to $J_3 > 0$ than for axisymmetric loadings corresponding to $J_3 < 0$, and as a consequence the strains at failure would be markedly different for these loadings (see also Fig. 5(a)). Since $\sigma_y^T/\sigma_y^C < 1$ and $\sigma_z^T/\sigma_z^C < 1$, the same trends are observed for axisymmetric loadings with the axial stress along the **y**-axis (TD) and **z**-axis (TT), respectively (see Fig. 5(b)-(c) and Eq. (18), and Eq. (19), respectively). Furthermore, the influence of the anisotropy of the material is clearly observed, the difference between the yield loci corresponding to stress states at $J_3 > 0$ and $J_3 < 0$, depending on the orientation of the applied axial stress (compare Fig 5(a)-(c)). Furthermore, unlike all existing criteria for porous solids, this criterion predicts that the response is strongly affected by couplings between the sign of the mean stress and that of $J_3$, the respective yield surfaces displaying a lack of symmetry with both the hydrostatic axis and deviatoric axis, respectively.

For example, the material's yield stress for purely tensile hydrostatic loadings is different than that under purely hydrostatic compressive loadings (see also Eq. (16)).

## 4. X-ray micro-tomography measurements on specimens with circular cross-section and comparison between model and data

Next, the predictive capabilities of the Cazacu and Stewart (2011) model are assessed. Using the values of the parameters of the model obtained from tensile test results on specimens with rectangular cross-section and compression tests (see Section 2.2), F.E. simulations of the response for uniaxial tension of specimens of circular cross-section were carried out. All the simulations were conducted with the same numerical values for the parameters involved in the model. Recall that the model parameters are those associated to the plastic properties of the material, namely the anisotropy coefficients $L_{ij}$, and strength differential parameter k (see Table 2), and the parameters associated to the isotropic hardening of the matrix material (Eq. (24)).

The average initial value of the void volume fraction of the specimens was estimated to be: $f_0$= 0.0001. The numerical values of the parameters involved in the void nucleation law (Eq.(14)) are: $f_N$ = 0.001, $S_N$ = 0.4, $\varepsilon_N$ = 0.9, $f_P$ = 0.001, $S_P$ = 250 MPa, $\sigma_P$ = 800 MPa. The elastic parameters values are: E = 110 GPa, $\nu$ = 0.3, where E is the Young modulus and $\nu$ is the Poisson coefficient. All the simulations were carried out using a user material subroutine (UMAT) that was developed for the constitutive model presented in Section 2 and implemented in the commercial implicit F.E. solver ABAQUS Standard (ABAQUS, 2009). A fully implicit integration algorithm was used for solving the governing equations.

Specifically, in the following the F.E. predictions are compared with the measured final cross-sections and X-ray computed micro-tomography (XCMT) measurements of porosity for both smooth and notched specimens.

**4.1. Experimental test results in uniaxial tension of axisymmetric smooth specimens and F.E. model predictions**

To further the understanding of the plasticity-damage couplings of the material, we performed additional uniaxial tension tests on smooth specimens of circular cross-section (radius of 3.18 mm) with axis along RD, TD, and at $15^o$, $45^o$ and at $75^o$ to the rolling direction. The specimen geometry and the F.E. meshes used in the simulations are shown in Fig. 6. Note that for the specimens oriented along the axes of orthotropy of the material (i.e. either RD or TD), only one-eighth of the specimen needs to be analyzed. The mesh used consisted of 8109 hexahedral elements (ABAQUS C3D8) (see Fig. 6 (b)). However, for off-axes specimens ($15^o$, $45^o$ and at $75^o$), symmetric boundary conditions cannot be applied, and the entire specimens were meshed, the F.E. mesh used in these cases consisting of 16355 hexahedral elements (see Fig. 6(c)).

To assess the predictive capabilities of the Stewart and Cazacu (2011) model, we first compare the predicted final cross-section with the one obtained experimentally in each test. In Fig. 7 are shown the photographs of the final cross-sections of the respective specimens on which are superposed the FE predictions obtained with the model (solid lines). It is worth noting that for all the loading orientations, the F.E. predictions are in good agreement with the experimental data. The Stewart and Cazacu (2011) porous model correctly captures the anisotropy in plastic deformation of the material, irrespective of the specimen orientation the initial circular cross-section becoming elliptical. Note also the marked difference in ellipticity of the final cross-sections of the specimens. The Stewart and Cazacu (2011) porous model predicts well the influence of the loading orientation on the shape of the final cross-section, the largest ellipticity being obtained for the RD specimen. Let define the ellipticity $e$, as

$$e = (a-b)/b, \tag{25}$$

where $a$ and $b$, are the major and minor axes of the respective deformed cross-section. For the RD specimen, the F.E. predicted ellipticity is of 28.3% against 30.4% obtained experimentally. For the specimens oriented along other directions, the ellipticity predicted by the model is of 25% for the $15^o$ specimen (against 26.5% experimentally), 22% for the $45^o$ specimen (against 21.7% experimentally), 21% for the $75^o$ specimen (against 24% experimentally) and 19.7% (against 22.7% experimentally). Thus, it can be concluded that the model predictions are in quantitative agreement with the data.

It is also worth noting the Stewart and Cazacu (2011) model captures fairly well the experimental axial load vs. displacement curves for all specimens. As an example, in Fig.8. is shown a comparison between the load-displacement curve obtained experimentally and the one according to the model for the RD specimen ( see Fig. 8(a) ) and for the $45^o$ specimen, respectively (see Fig. 8(b)).

It is worth recalling that all the simulations were conducted with the same set of values for the model parameters which were identified based on compression data and tension data obtained on flat specimens. Thus, the simulation results presented in Fig. 7-8 are fully predictive of the plastic behavior of the material. Next, we will assess the capabilities of the model to predict porosity evolution in HCP-titanium.

## 4.2. XCMT measurements of damage for axisymmetric smooth RD specimen

With the availability of High-Resolution X-ray Computed Tomography (XRCT), in the past decade it has become possible the quantification of damage in the bulk of a given material. To have information on damage in the material under uniaxial tension, in the present work, images obtained by X-ray computed microtomography (using Xradia X-ray microscope, VersaXRM-500) were taken at the necking region of an axisymmetric smooth RD specimen. The XCMT scan was done after the specimen was subjected to uniaxial tension up to an axial displacement of 5.52 mm. It is worth noting that the XCMT scan was taken very close to failure. In Fig.9(a) are shown the reconstructed 2-D views in the (TD, TT) plane, and in the (RD, TD) plane of the specimen, respectively. For comparison purposes, in Fig. 9(b) are shown the respective views extracted from an XCMT scan of a copper specimen, which was taken at the same axial displacement. It is very important to note that for the same axial displacement, the copper material is very damaged while titanium material shows very little damage. While a large hole/crack is seen in the middle of the copper specimen, almost no damage is observed in the HCP–Ti specimen. It means that for uniaxial tensile loading the rate of damage growth is slower in Ti than in Cu. Thus, these XCMT observations confirm the conclusions of the preliminary theoretical study of Revil-Baudard and Cazacu (2013) on the influence of the tension-compression asymmetry on damage growth rate and damage distribution in a round tensile specimen. Using the isotropic form of Stewart and Cazacu (2011) model (i.e. Eq. (20)), in this study it was shown that for materials for which the flow stress in uniaxial compression is larger than the flow stress in uniaxial tension damage is delayed as compared to materials for which the matrix plastic behavior is governed by the von Mises criterion. The observed drastic difference in damage evolution (see Fig.9) between HCP- titanium, which is harder in compression than in tension (see Fig. 4), and the FCC copper material, which does not display tension-compression asymmetry and its plastic behavior can be approximated with the von Mises criterion validates the main conclusion of Revil-Baudard and Cazacu (2013) study, the rate of void growth being much lower for Ti than for copper. Furthermore, Fig. 9 clearly shows that there is a very strong coupling between the specificities of the plastic deformation and porosity/damage evolution.

**4.3. Comparison between model prediction and XCMT measurements for an axisymmetric smooth RD specimen**

In order to assess the predictive capabilities of the Steward and Cazacu (2011) porous model in terms of damage evolution in HCP-Ti, the XCMT data were compared with the F.E. isocontours of the void volume fraction corresponding to the same axial displacement (see Fig.10). Specifically, the F.E. predictions are superposed on the different views obtained by XCMT, namely on the experimental cross-section, the experimental (RD, TD) section, and the experimental (RD,ND). It is worth noting that the Stewart and Cazacu (2011) model predicts diffuse porosity, in the minimal cross-section the maximum void volume fraction predicted being of 0.22% (see Fig.10a). Experimentally, the void volume fraction observed in same cross-section is of 0.25%. Likewise, for the other views (Fig. 11(b) and Fig. 11(c), respectively), most of the voids observed by XCMT are inside the region of maximum void volume fraction predicted by the model. The analysis of the XCMT data was done using ImageJ, which is a public domain image processing program developed at the National Institutes of Health (see Rasband, 2014; Schneider et al., 2012; Abramoff et al., 2004).

Furthermore, in Fig.10, the F.E. predictions of the geometry of the specimen for the respective axial displacement (red points) are superposed on the different cut views. It is to be noted that the Stewart and Cazacu (2011) model correctly predicts the section geometry in all the planes.

In addition, in Fig. 11 are shown the F.E. predictions of the isocontours of the void volume fraction predicted for the same specimen corresponding to an axial displacement of 5mm, 5.6 mm, and 5.85 mm, respectively. It is shown that for an axisymmetric smooth specimen, according to the model damage initiates at the center (see Fig. 11(a)), and shifts toward the outside surface of the specimen as the axial displacement increases (see Fig. 11(b)-(c)).

## 4.4. In-situ XCMT measurements of damage evolution for an axisymmetric notched RD specimen of Ti and comparison with model predictions

In the previous section, it was shown that Cazacu and Stewart (2011) porous model predicts correctly the level of damage as well as plasticity damage couplings couplings under uniaxial tensile loading of an axisymmetric smooth specimen. The comparison between model and data were done for a fixed level of the imposed axial displacement. Theoretically, it was shown that damage evolution in the $\alpha$-Ti has very unusual and unexpected characteristics (see Fig. 11).

Next, we will assess the predictive capabilities of the model to capture the influence of the stress-state (stress triaxiality) on damage and its evolution. For this purpose, an uniaxial tension test up to fracture was conducted on a notched axisymmetric RD specimen. The notch radius was of 0.51mm while the specimen cross-section radius was of 1.27 mm (see Fig. 12(a)). The load-displacement curve is shown in Fig. 13 in comparison with the Stewart and Cazacu (2011) FE model predictions. Given that the specimen has its axis along an orthotropy axis of the material (RD), only one-eighth of the specimen needs to be meshed. The mesh used consists of 6803 hexahedral elements (ABAQUS C3D8) (see Fig. 12(b)). On Fig. 13 are also shown for comparison the load vs. displacement curves obtained experimentally and predicted by the model for the smooth RD specimen. It is clearly seen that the model describes well that the presence of the notch induces a much softer response as compared to the smooth specimen.

To further provide insights into the porosity evolution in the material and verify the unusual damage characteristics revealed by the model, in-situ XCMT tensile tests on RD specimens of the same notch radius were conducted at Wright Patterson Air Force Laboratory.

Due to the specificity of the in-situ XCMT testing capabilities, the load-displacement curve cannot be obtained directly. However, for each XCMT scan, knowing the resolution of the X-ray microscope (1 pixel is equal to 3.2516 μm), it was possible to deduce the displacement between the extremities of the notch. In Fig. 14 are shown views obtained from scans taken at notch displacement of 0.24 mm, 0.73mm, 1.02 mm and 1.20 mm, respectively. To assess the effect of anisotropy on the plastic and damage response, at each individual level of displacement, the cross-sections in the (TD, ND) plane), in the (RD, TD) plane, and in the (RD, ND) plane are reported (see Fig. 14). Model predictions of the porosity isocontours corresponding to various cross-sections and several levels of axial displacements are presented in Fig. 15- Fig. 17.

A close examination of the scans shown in Fig.14 for the notched specimen reveals that damage initiates at the surface of the specimen. Thus, the unusual damage evolution characteristics predicted by the model (see Fig. 16; Fig. 17) are confirmed experimentally. Furthermore, it is demonstrated that the location of the zone where damage initiates strongly depends on the specimen geometry. In other words, for the same global loading (i.e. uniaxial tension), the local stress state, which depends on the specimen geometry is different, which in turn triggers a different damage initiation site, and a markedly different damage evolution. Indeed, the model predicts that for a smooth axisymmetric specimen, damage initiates at the center of the specimen (see Fig.11), and the level of damage close to failure is very low (see also the scan of Fig. 9(a) which confirms the trends). On the other hand, for the notched specimen the model predicts that damage initiates at the outer surface of the specimen, and further grow from the outer surface to the center of the specimen (see Fig. 16-Fig. 17) , which corroborates with the in-situ XCMT data of Fig.14 for the notched specimen. Thus, it is clearly demonstrated that damage initiation and evolution in the titanium material strongly depends on the stress-state.

It is very important to note that the model predicts damage anisotropy. The predictions of the isocontours of porosity in the cross-section corresponding to different levels of notch displacement (e.g. see Fig. 15(d)) indicates that the void volume fraction is larger along TD (i.e. the small axis of the deformed elliptical cross-section) than along ND (i.e. the long axis of the elliptical cross-section). Moreover, comparison of the XCMT (RD, TD) sections with the (RD,ND) sections (see Fig. 14), clearly shows that the surface of the specimen along the TD axis is more damaged than the surface along the ND axis. The prediction of the Stewart and Cazacu (2011) porous model corroborates with these experimental observations (see Fig. 16 and Fig. 17).

To further demonstrate that there is a good agreement between the predictions of the Stewart and Cazacu (2011) model in terms of plasticity-damage couplings in Fig 16-17 are also superposed on the XCMT scans the F.E. predictions of the specimen profiles (red points) corresponding to notch displacements of 0.73mm and 1.2 mm, respectively. It should be noted that the model correctly captures the change in the geometry of the specimen (anisotropy in plastic deformation). Since damage is driven by the plastic deformation, and the plastic anisotropy is correctly described, the model also correctly predicts the location of the zones of maximum damage in each plane.

## 5. Summary and Conclusions

In this paper, results of an experimental study on the room-temperature quasi-static plastic deformation and damage of a high-purity, polycrystalline, HCP-titanium material were presented. To quantify the plastic anisotropy and the tension-compression symmetry of the material, first monotonic uniaxial compression and tension tests were carried out. The test results indicated that the material is orthotropic. Although the true stress-strain curves for in-plane specimens are close, there is a marked difference between the yield stress in the normal plate direction and the in-plane directions. Irrespective of the loading orientation, the material displays strength-differential effects (harder in compression than in tension). Digital image correlation techniques have been used to determine the strain fields (axial and width strains) in the gage zone. To characterize the evolution of anisotropy with accumulated plastic strain, for each specimen orientation, Lankford coefficients (r-values) were determined for several levels of plastic strain. Irrespective of the strain level, the largest r-value is along RD.

Stewart and Cazacu (2011) anisotropic elastic/plastic damage model was applied to describe the behavior of the material studied. All material parameters involved in the Stewart and Cazacu (2011) damage model have a clear physical significance, being related to plastic properties. Specifically, these material parameters were identified based on the experimental data obtained in a few mechanical characterization tests (uniaxial tension tests on specimens of rectangular cross-section, and uniaxial compression tests).

In addition to the mechanical characterization tests on flat tensile specimens, uniaxial tensile tests on cylindrical specimens with circular cross-section were also conducted. To further the understanding of the plasticity-damage couplings of the material, and assess the predictive capabilities of the model, both ex-situ and in-situ XCMT measurements were done on the specimen oriented along the rolling direction. Specifically, the F.E. predictions obtained using the same set of parameters previously identified were compared with the measured final cross-sections and XCMT measurements of porosity for both smooth and notched specimens.

The XCMT measurements on axisymmetric smooth specimens reveal the drastic difference in damage evolution between α-titanium, which is harder in compression than in tension and acopper material that does not display tension-compression asymmetry. For the same axial displacement, the copper material is very damaged while titanium material shows very little damage. While a large hole/crack is seen in the middle of the copper specimen, almost no damage is observed in the HCP–Ti specimen.

It was shown that the model predicts correctly both the anisotropy of plastic deformation (i.e. accurate predictions of the final geometry of the specimens irrespective of the loading orientation), and damage in Ti. Specifically, for a smooth axisymmetric specimen subject to uniaxial tension, damage initiates at the center of the specimen and is diffuse; the level of damage close to failure is very low. On the other hand, for a notched specimen subject to uniaxial tension the model predicts that damage initiates at the outer surface of the specimen, and further grows from the outer surface to the center of the specimen, which corroborates with the in-situ XCMT data.

# References

ABAQUS., 2009. User's Manual for Version 6.8, vol. I–V. Dassault Systemes Simulia Corp., Providence, RI.

Abramoff, M.D., Magalhaes, P.J., Ram, S.J., 2004. Image Processing with ImageJ. Biophotonics International, 11(7), 36-42.

Argon, A. S., Im, J., and Safoglu, R., 1975. Cavity formation from inclusions in ductile fracture. *Metallurgical Transactions A*, *6*(4), 825-837.

Barlat, F., Richmond, O., 1987. Prediction of tricomponent plane Stress yield surfaces and associated flow and failure behavior of strongly textured F.C.C. polycrystalline sheets. Mat. Sci. Eng. 95, 15-29.

Barlat, F., Cazacu, O., Zyczkowski, M., Banabic, D. and Yoon, J.W. "Yield surface plasticity and anisotropy in sheet metals" in: *Continuum Scale Simulation of Engineering Materials, Fundamentals - Microstructures - Process Applications*, Weinheim,Wiley-VCH Verlag Berlin GmbH, 145, 2004.

Banerjee, D., Williams , J.C., 2013. Perspectives on titanium science and technology, Acta Mater. 61, 844-879.

Benhenni, N., Bouvier, S., Brenner, R., Chauveau, T., Bacroix, B., 2013. Micromechanical modelling of monotonic loading of CP-a titanium: correlation between macroscopic and microscopic behavior. Mat. Sci. Eng. A., 573, 222.

Bouvier, S., Benmhenni, N., Tirry, W., Gregory, F., Nixon, M.E. , Cazacu, O., Rabet, L., 2012. Hardening in relation with microstructure evolution of high-purity alpha-titanium deformed under monotonic and cyclic loading at room temperature, Mat. Sci. Eng. A 535, 12-21.

3
Cazacu, O., Barlat, F., 2004. A criterion for description of anisotropy and yield differential effects in pressure-insensitive metals. Int. J. Plasticity, 20(11), 2027-2045.

Cazacu, O., Plunkett, B., Barlat, F., 2006. Orthotropic yield criterion for hexagonal closed packed metals. Int. J. Plasticity, 22(7), 1171-1194.

Cazacu, O., Stewart, J. B., 2009. Analytic plastic potential for porous aggregates with matrix exhibiting tension–compression asymmetry. J. Mech. Phys. Solids, 57(2), 325-341.

Cazacu, O., Revil-Baudard, B., Lebensohn, R. A., Garajeu, M., 2013. On the combined effects of pressure and third-invariant on yielding of porous solids with von Mises matrix. J. Appl. Mech. 80 (6), 064501-064501-5.

Cazacu, O., Revil-Baudard, B., Chandola N., Kondo, D., 2014. Analytical criterion for porous solids with Tresca matrix accounting for combined effects of the mean stress and third-invariant of the stress deviator. Int. J. Solids Structures 51, 861-874.

Chu, C. C., and Needleman, A., 1980. Void nucleation effects in biaxially stretched sheets. Journal of Engineering Materials and Technology, Transactions of the ASME, 102(3), 249-256.

Chun, Y.B, Yu S.L, Semiatin S.L., Hwang, S.K., 2015. Effect of deformation twinning on microstructure and texture evolution during cold rolling of CP-titanium. Mater. Sci. Eng A 2005;398:209–19.

Gilles, G., Hammami, W., Libertiaux, V. Cazacu, O. Yoon, J.H., Kuwabara, T., Habraken, A.M., Duchêne, L., 2011. Experimental characterization and elasto-plastic modeling of the quasi-static mechanical response of TA–6 V at room temperature. Int. J. Solids Struct. 48, 1277-1289.

Gilles G., Habraken A.M., Cazacu O., Balan T. and Duchêne L., 2013. Experimental and numerical study of TA6V mechanical behavior under different quasi-static strain paths at room temperature. AIP Conference Proceedings, 1532:651-656.

Gurland, J. (1972). Observations on the fracture of cementite particles in a spheroidized 1.05% C steel deformed at room temperature. *Acta Metallurgica*, *20*(5), 735-741.

Hill, R. 1948, A theory of the yielding and plastic flow of anisotropic materials. Proc. Royal. Soc. Lond. A 193, 281-297.

Hill, R., 1967. The essential structure of constitutive laws for metal composites and polycrystals. J. Mech. Phys. Solids. 15, 79-95.

Huez, J., Feugas, X., Helbert, A. L., Guillot, I., and Clavel, M., 1998. Damage process in Commercially Pure α-Titanium Alloy without (Ti40) and with (Ti40-H) Hydrides. Metall. Mater. Trans. A., 29, 1615.

Knezevic, M., Lebensohn, R. A., Cazacu, O., Revil-Baudard, B., Proust, G., Vogel, S. C., Nixon, M. E., 2013. Modeling bending of α-titanium with embedded polycrystal plasticity in implicit finite elements. Mat. Sci. Eng. A 564, 116-126.

Kuwabara T, Katami C, Kikuchi M, Shindo T, Ohwue T. Cup drawing of pure titanium sheet-finite element analysis and experimental validation. In: 7th Int. Conference on Numerical Methods in Industrial Forming Processes, 18–20 June, Toyohashi, Japan; 2001, 781–7.

Lebensohn, R.A., Cazacu, O., 2012. Effect of single-crystal plastic deformation mechanisms on the dilatational plastic response of porous polycrystals. Int. J. Solids Struct., 49, 3838-3852.

Mandel J., 1972. Plasticité classique et viscoplasticité. CISM Courses and Lectures, Vol.97, International Center for Mechanical Sciences, Springer-Verlag, Wien-New York.

Nixon, M. E., Lebensohn, R. A., Cazacu, O., Liu, C. 2010a. Experimental and finite-element analysis of the anisotropic response of high-purity α-titanium in bending. Acta Mater., 58, 5759-5767.

Nixon, M. E., Cazacu, O., Lebensohn, R.A., 2010b. Anisotropic response of high-purity α-titanium: experimental characterization and constitutive modeling. Int. J. Plasticity 26, 510–532.

Rasband, W.S., ImageJ, U. S. National Institutes of Health, Bethesda, Maryland, USA, http://imagej.nih.gov/ij/, 1997-2014.

Revil-Baudard, B., and Cazacu, O., 2013. On the effect of the matrix tension–compression asymmetry on damage evolution in porous plastic solids. European Journal of Mechanics-A/Solids, 37, 35-44.

Revil-Baudard, B., Cazacu, O, Chandola, N., Barlat, F., 2014. Correlation between Swift effects and tension-compression asymmetry in various polycrystalline materials, J. Mech. Phys. Solids., 70, 104-115.

Revil-Baudard, B., Cazacu, O, Flater, P., Kleiser, G., 2015. Plastic deformation of high-purity α-titanium: model development and validation using the Taylor cylinder impact test, Mechanics of Materials, 80, 2015, 264–275.

Rice, J.R., Tracey, D.M., 1969. On the ductile enlargement of voids in triaxial stress fields. J. Mech. Phys. Solids 17, 201-217.

Schneider, C.A., Rasband, W.S., Eliceiri, K.W., 2012. NIH Image to ImageJ: 25 years of image analysis. Nature Methods, 9, 671-675.

Stewart, J. B., and Cazacu, O., 2011. Analytical yield criterion for an anisotropic material containing spherical voids and exhibiting tension–compression asymmetry. Int. J. Solids Struct., 48(2), 357-373.

## Appendix A: Components of the 4$^{th}$ order anisotropic tensor B

Using Voigt notations, the fourth-order anisotropic tensor **L** involved in the expression of the criterion given by Eq. (1) is represented in the reference system associated with orthotropy by:

$$\mathbf{L} = \begin{bmatrix} L_{11} & L_{12} & L_{13} & & & \\ L_{12} & L_{22} & L_{23} & & & \\ L_{13} & L_{23} & L_{33} & & & \\ & & & L_{44} & & \\ & & & & L_{55} & \\ & & & & & L_{66} \end{bmatrix} \quad (A.1)$$

The symmetric tensor **C** = **LK** is invertible i.e. there exist a 4$^{th}$ order symmetric tensor **B** such that

$$\mathbf{B\,C} = \mathbf{K} \quad (A.2)$$

where **K** is the 4$^{th}$ order symmetric deviatoric unit tensor. In Voigt notations

$$\mathbf{K} = \begin{bmatrix} 2/3 & -1/3 & -1/3 & 0 & 0 & 0 \\ -1/3 & 2/3 & -1/3 & 0 & 0 & 0 \\ -1/3 & -1/3 & 2/3 & 0 & 0 & 0 \\ 0 & 0 & 0 & 1 & 0 & 0 \\ 0 & 0 & 0 & 0 & 1 & 0 \\ 0 & 0 & 0 & 0 & 0 & 1 \end{bmatrix} \quad (A.3)$$

Thus, the components of the tensor **B** (Voigt notations) are:

$$B_{12} = \frac{1}{3} \cdot \frac{2(C_{32} - C_{12}) + (C_{31} - C_{11})}{(C_{21} - C_{11})(C_{32} - C_{12}) - (C_{22} - C_{12})(C_{31} - C_{11})}$$

$$B_{13} = \frac{1}{3} \cdot \frac{(C_{11} - C_{21}) + 2(-C_{22} + C_{12})}{(C_{21} - C_{11})(C_{32} - C_{12}) - (C_{22} - C_{12})(C_{31} - C_{11})} \quad (A.4)$$

$$B_{11} = -(B_{12} + B_{13})$$

$$B_{21} = B_{12}$$

$$B_{23} = \frac{1}{3} \cdot \frac{2(C_{11} - C_{21}) + (C_{12} - C_{22})}{(C_{11} - C_{21})(C_{32} - C_{22}) - (-C_{22} + C_{12})(C_{31} - C_{21})}$$

$$B_{22} = -(B_{21} + B_{23})$$

$$B_{31} = B_{13}$$

$$B_{32} = B_{23}$$

$$B_{33} = -(B_{31} + B_{23})$$

$$B_{44} = \frac{1}{C_{44}}$$

$$B_{55} = \frac{1}{C_{55}}$$

$$B_{66} = \frac{1}{C_{66}}.$$

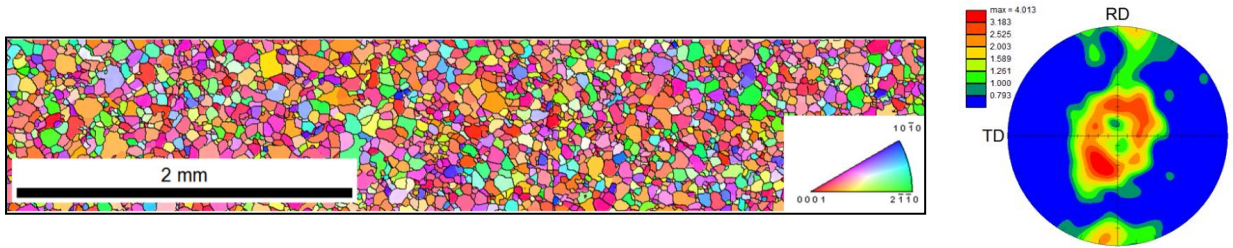

**Fig**. 1: Orientation map and basal {0001} pole figure showing the initial texture of the α-titanium plate material, measured using Electron Back Scattered Diffraction (EBSD). The rolling directions (RD) and transverse direction (TD) are in the plane of the plate.

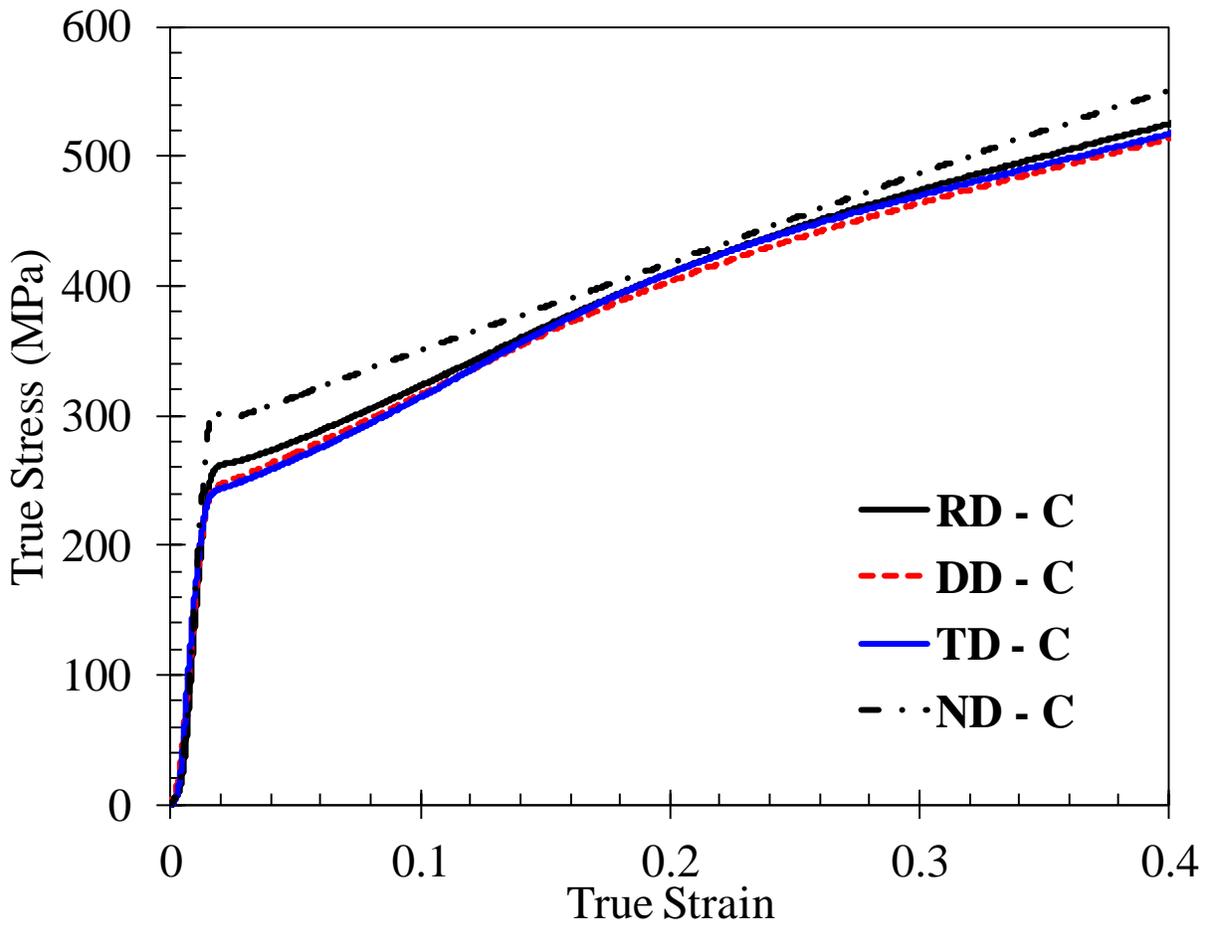

**Fig**. 2: Experimental quasi-static uniaxial compression stress-strain curves along several in-plane orientations θ measured with respect to the rolling direction (RD) (i.e. θ= 0° (RD), 45° (DD), and 90° (TD), and in the normal direction (ND) of the HCP-titanium plate.

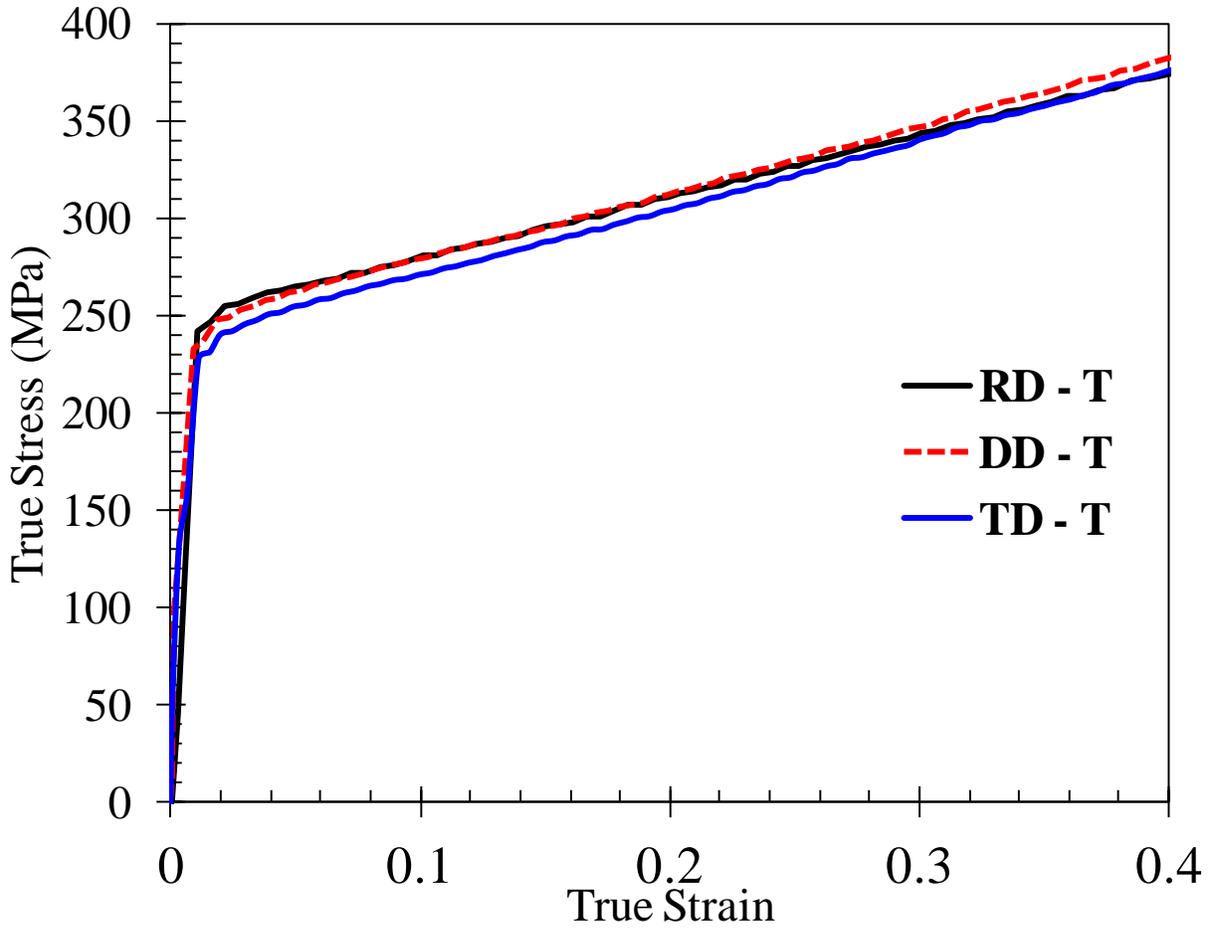

**Fig. 3**: Experimental quasi-static uniaxial tension stress-strain curves along several in-plane orientations θ measured with respect to the rolling direction (RD):. θ= 0° (RD), 45° (DD), and 90° (TD)) of the HCP-titanium plate.

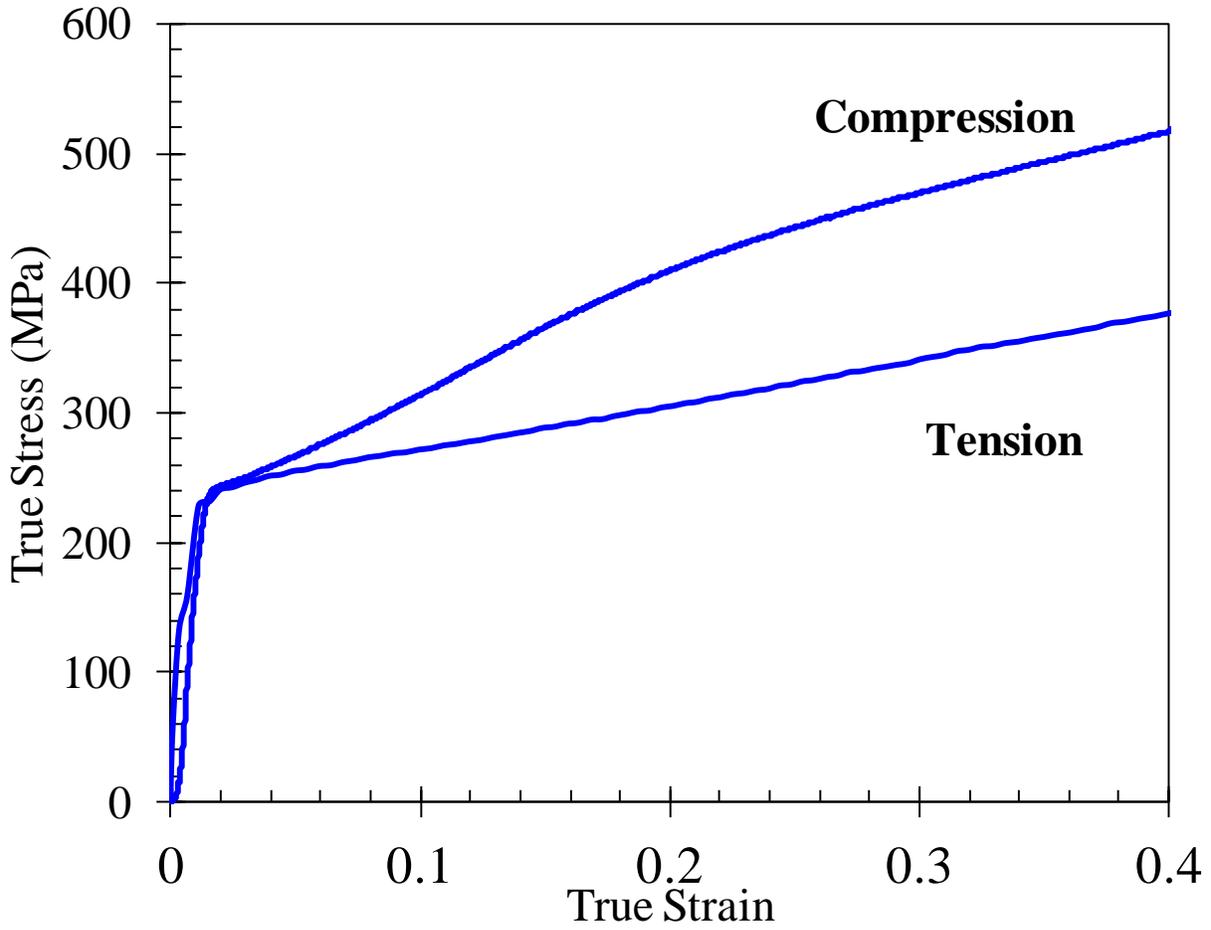

**Fig.** 4: Experimental evidence of the tension-compression asymmetry along the transverse direction (TD) in the plane of the HCP-titanium plate. Note that the material is harder in compression than in tension.

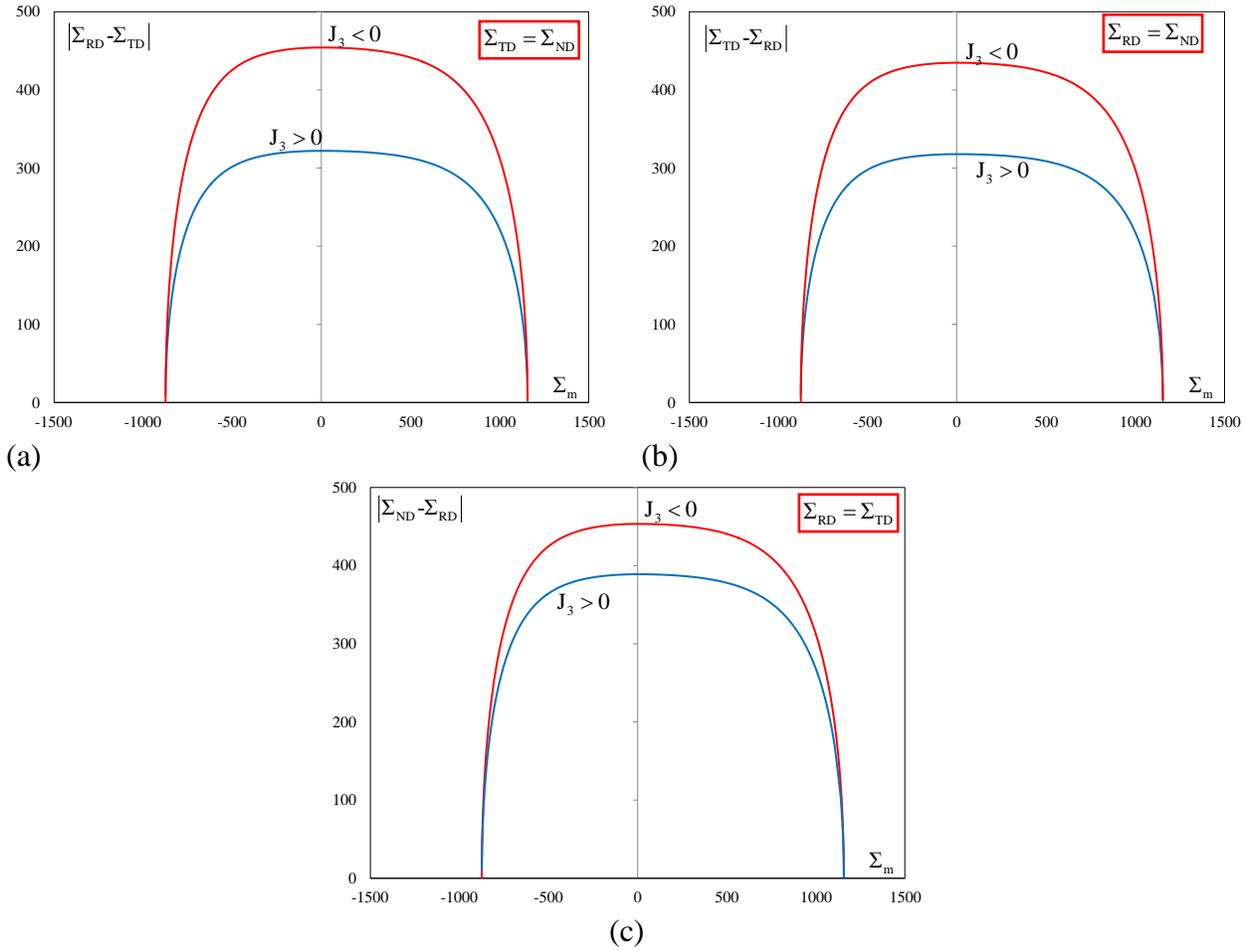

**Fig**. 5: Stewart and Cazacu (2011) surfaces for the porous anisotropic HCP-titanium corresponding to a porosity $f = 0.01$ and plastic strain of 0.25 for axisymmetric loadings with axial stress oriented either along RD, TD or ND :(a) axisymmetric loading along RD-axis such that $\Sigma_{RD} > \Sigma_{ND}=\Sigma_{TD}$ ($J_3 > 0$) and $\Sigma_{RD} < \Sigma_{ND} = \Sigma_{TD}$ ($J_3 < 0$); (b) axisymmetric loading along TD-axis for $\Sigma_{TD} > \Sigma_{RD} = \Sigma_{ND}$ ($J_3 > 0$) and $\Sigma_{TD} < \Sigma_{RD} = \Sigma_{ND}$ ($J_3 < 0$); (c) axisymmetric loading along ND-axis for $\Sigma_{ND} > \Sigma_{RD} = \Sigma_{TD}$ ($J_3 > 0$) and $\Sigma_{ND} < \Sigma_{RD} = \Sigma_{TD}$ ($J_3 < 0$), respectively.

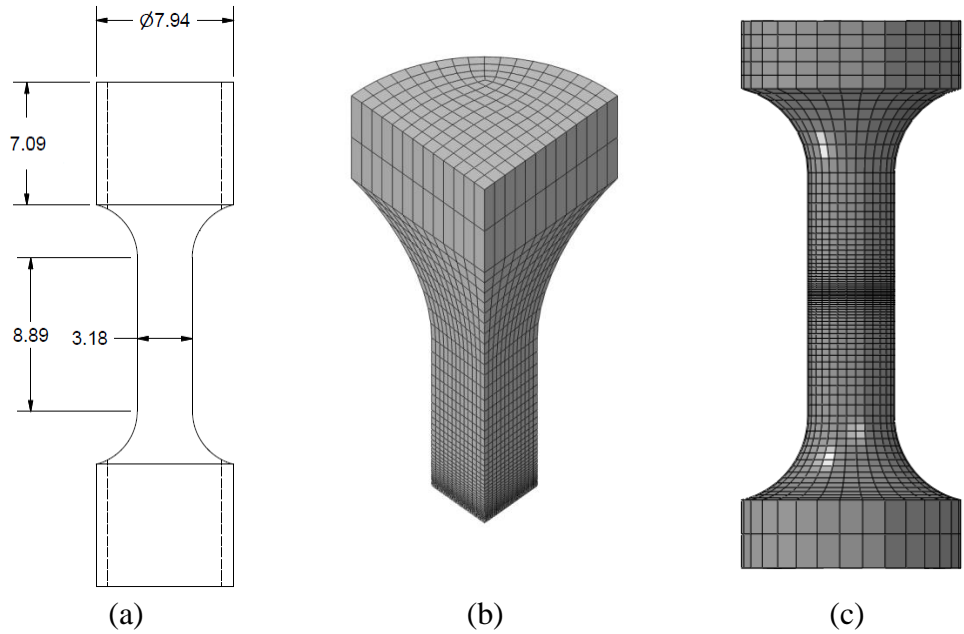

**Fig**. 6: (a) Geometries of the axisymmetric smooth tensile specimen; (b) F.E. meshes used for the specimens with axes along the axes of orthotropy: rolling direction (RD) and transverse direction (TD); (c) F.E. meshes used for the off-axis specimens i.e. the 15°, 45° and at 75° to the rolling direction (RD). Dimensions are in mm.

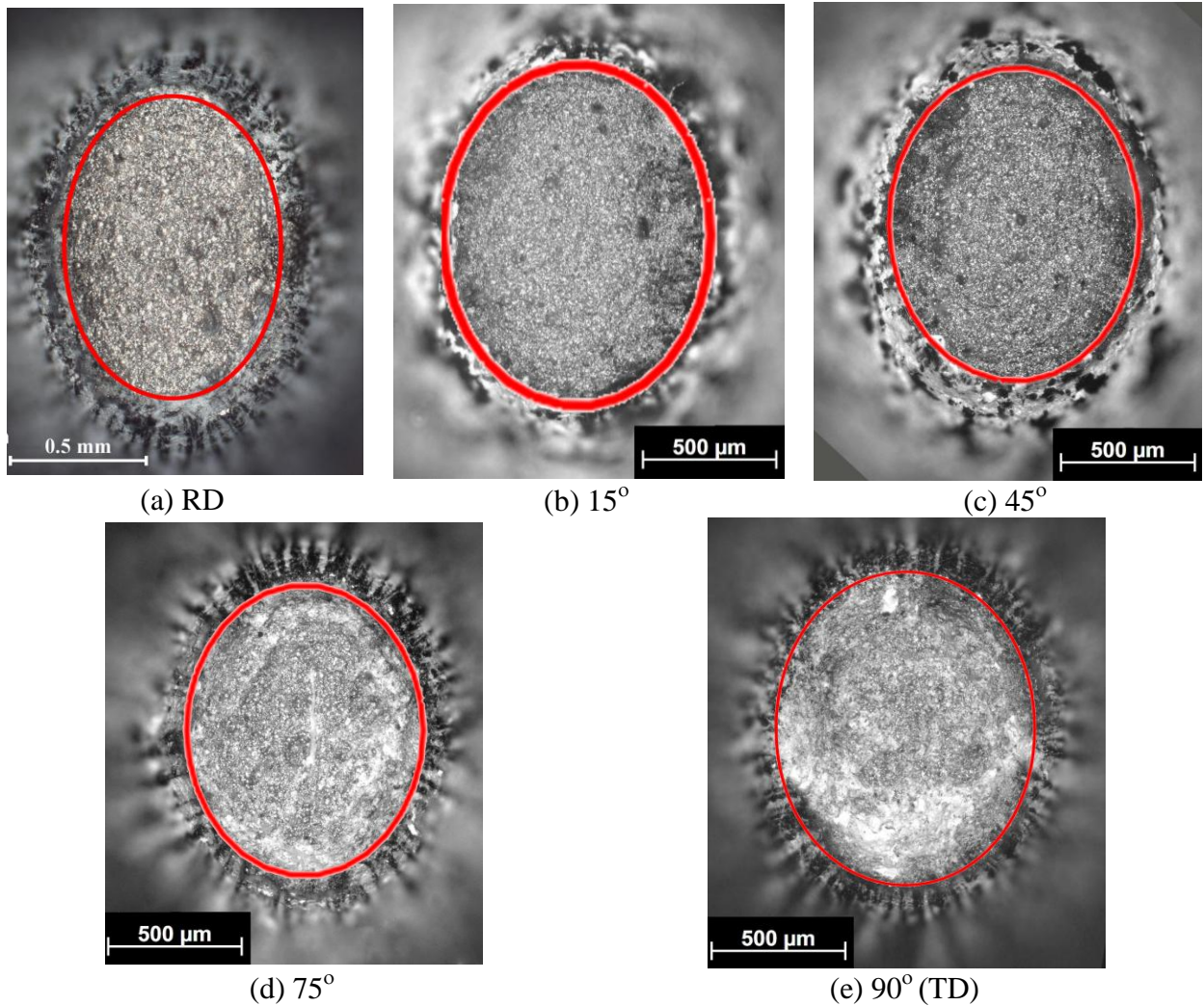

**Fig**. 7: Comparison between experiments and F.E. predictions according to the Stewart and Cazacu (2011) model for the final cross-sections of axisymmetric tensile specimens: (a) specimen with axis along the rolling direction (RD); (b) specimen with axis at $15\,^{\circ}$ to RD; (c) specimen with axis at $45\,^{\circ}$ to RD; (d) specimen with axis at $75\,^{\circ}$ to RD; (e) specimen with axis at $90\,^{\circ}$ to RD, respectively. The F.E. predictions, represented as solid lines, are superposed on the photographs of the deformed specimens.

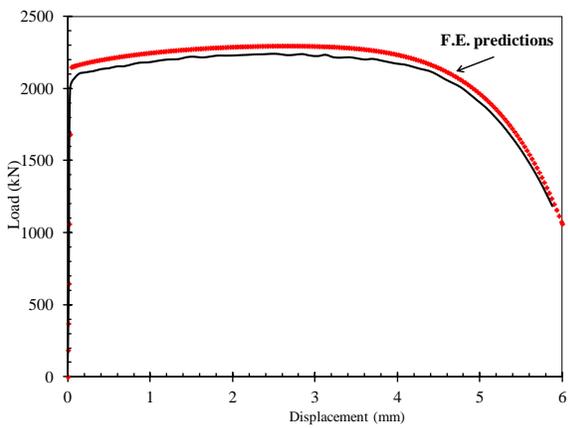 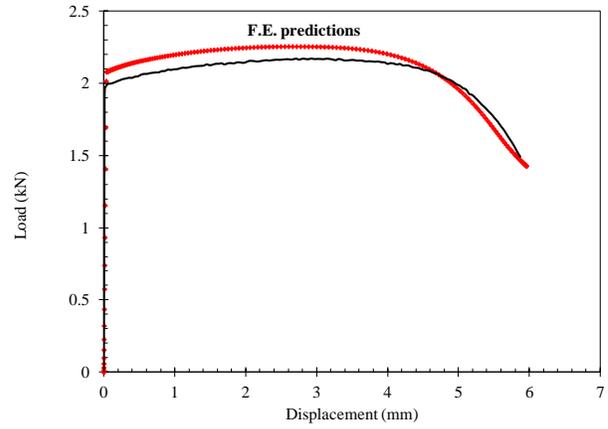

(a) RD  (b) 45º to the RD

**Fig**. 8 Comparison between experimental data and F.E. predictions of the load vs. axial displacemet curves according to the Stewart and Cazacu (2011) model for uniaxial tension of smooth axisymmetric specimens: (a) specimen with axis along the rolling direction (RD); (b) specimen with axis at 45º to RD, respectively.

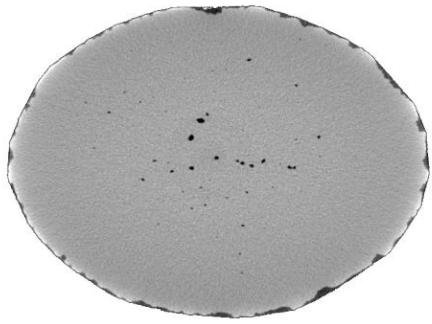 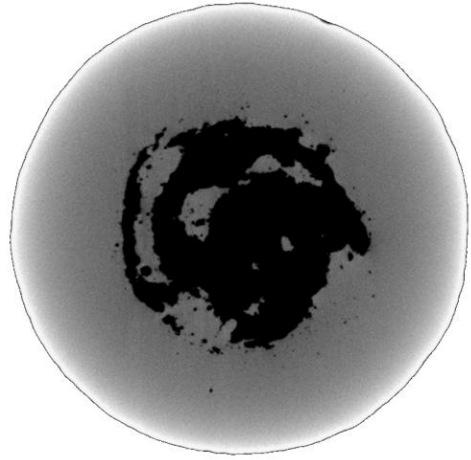

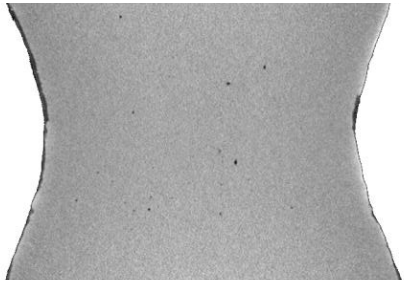 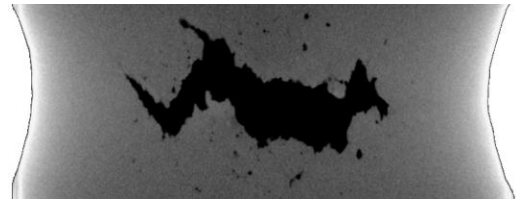

(a) HCP-titanium       (b) Copper

**Fig**. 9: X-ray computed micro-tomography (XCMT) scans close to failure: (a) HCP-titanium material studied ; (b) isotropic copper material.

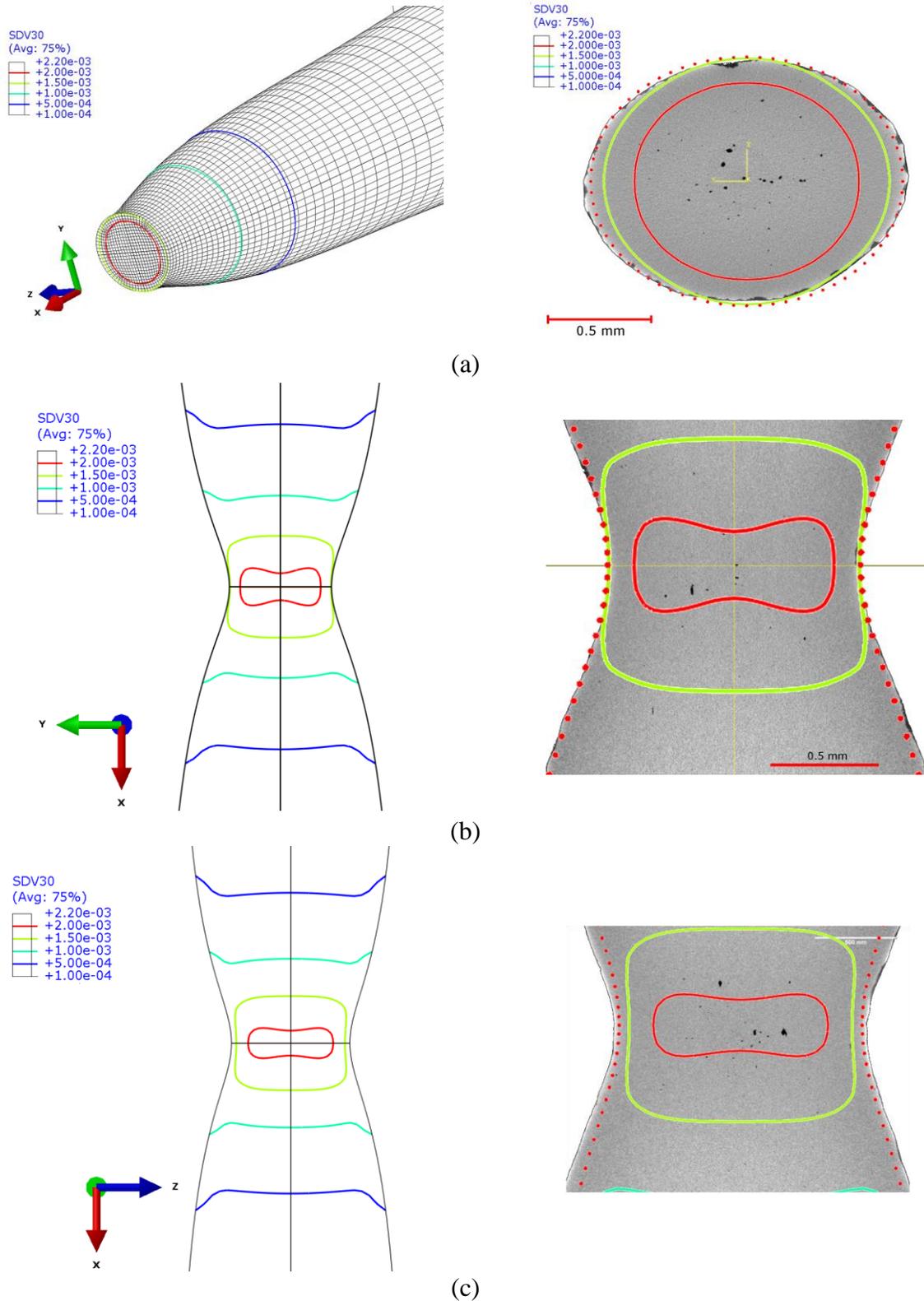

**Fig**.10. Comparison between the F.E. cross-sections and isocontours of void volume fraction of a smooth axisymmetric specimen of HCP-titanium subjected to uniaxial tension along the rolling direction (RD) according to the Stewart and Cazacu (2011) model and XCMT data for an axial

notch displacement of 5.52 mm: (a) deformed specimen cross-section; (b) (RD-TD) section of the deformed specimen; (c) (RD-ND) section of the deformed specimen. The rolling, transverse, and normal directions are denoted as RD, TD, ND.

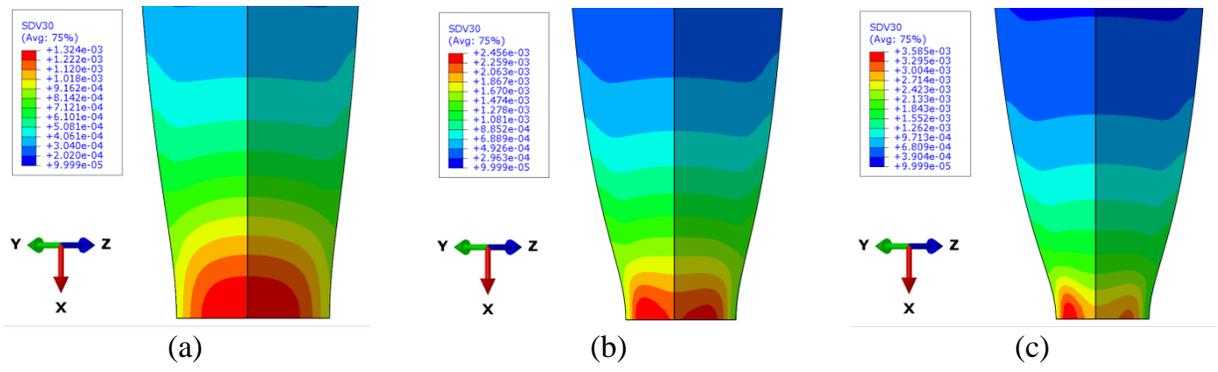

**Fig**.11: Isocontours of the void volume fraction according to the Stewart and Cazacu (2011) model for an axisymmetric smooth specimen of HCP-titanium subjected to uniaxial tension along the rolling direction (RD) specimen for an axial displacement of: (a) 5 mm; (b) 5.6 mm; (c) 5.85 mm. Note that the location of the zone of maximun void volume fraction shifts away from the center, at higher levels of axial displacement.

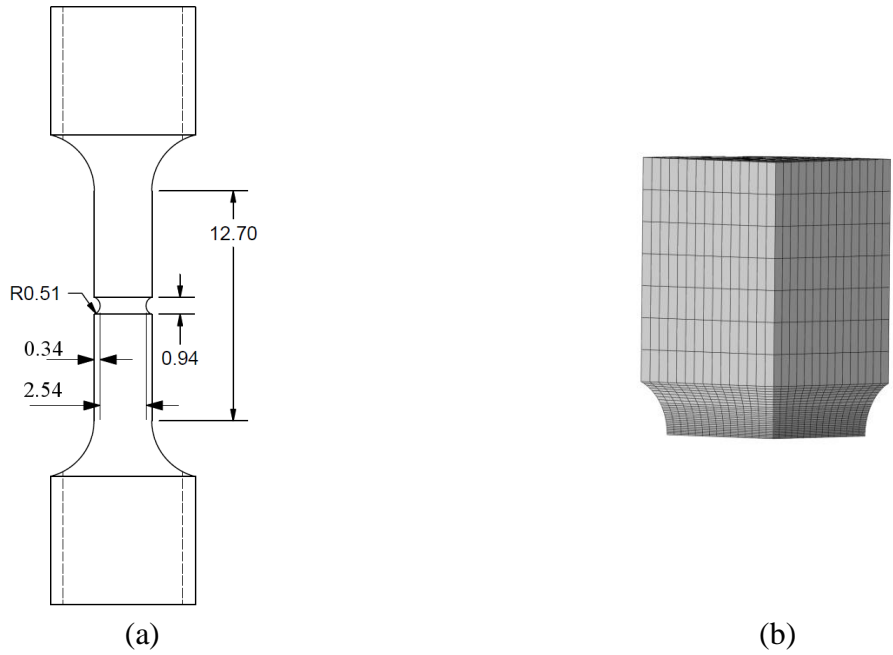

**Fig**. 12: (a) Geometry of the notched specimen used for unixial tension along the rolling direction (RD) and (b) F.E. mesh. Dimensions are in mm.

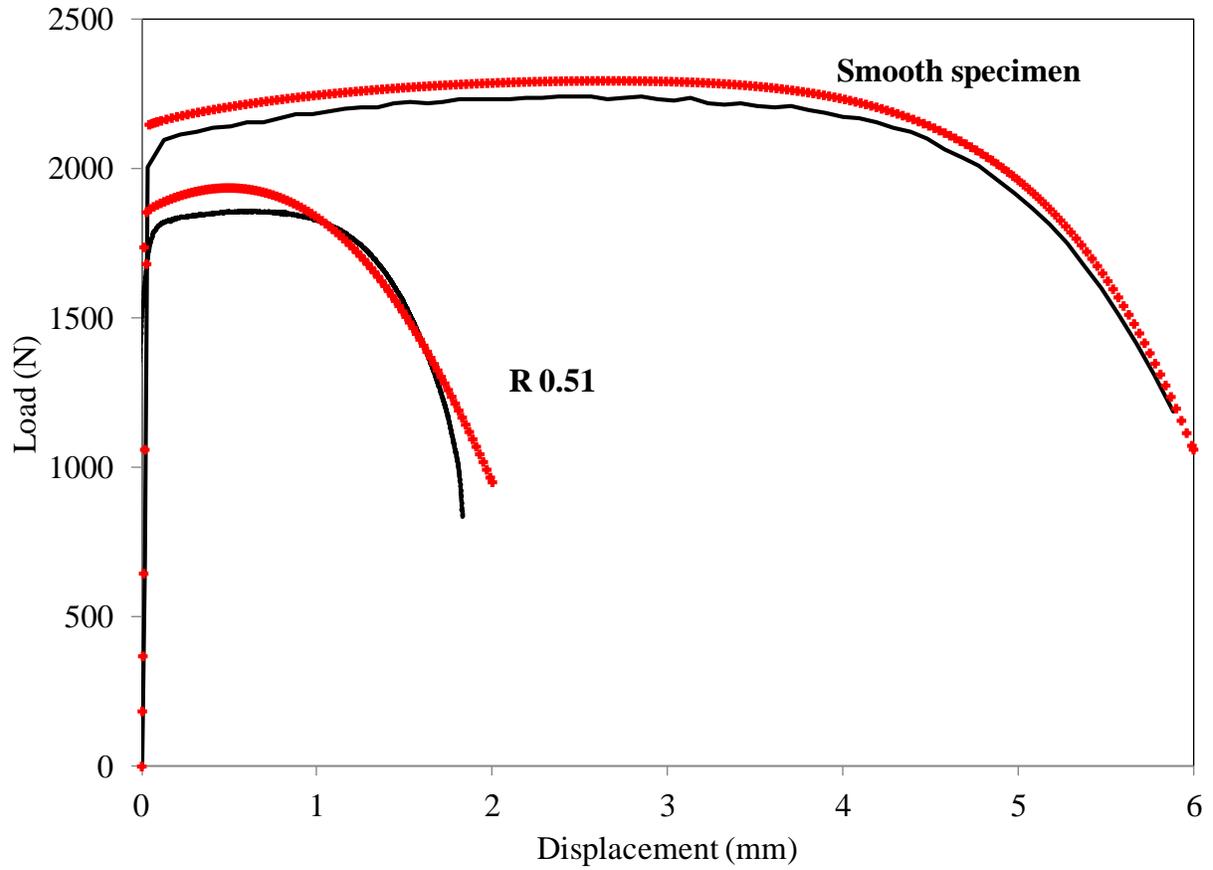

**Fig**. 13. Effect of the notch geometry on the load vs. displacement response for HCP-titanium according to the Stewart and Cazacu (2011) model and experimental data in uniaxial tension along the rolling direction (RD).

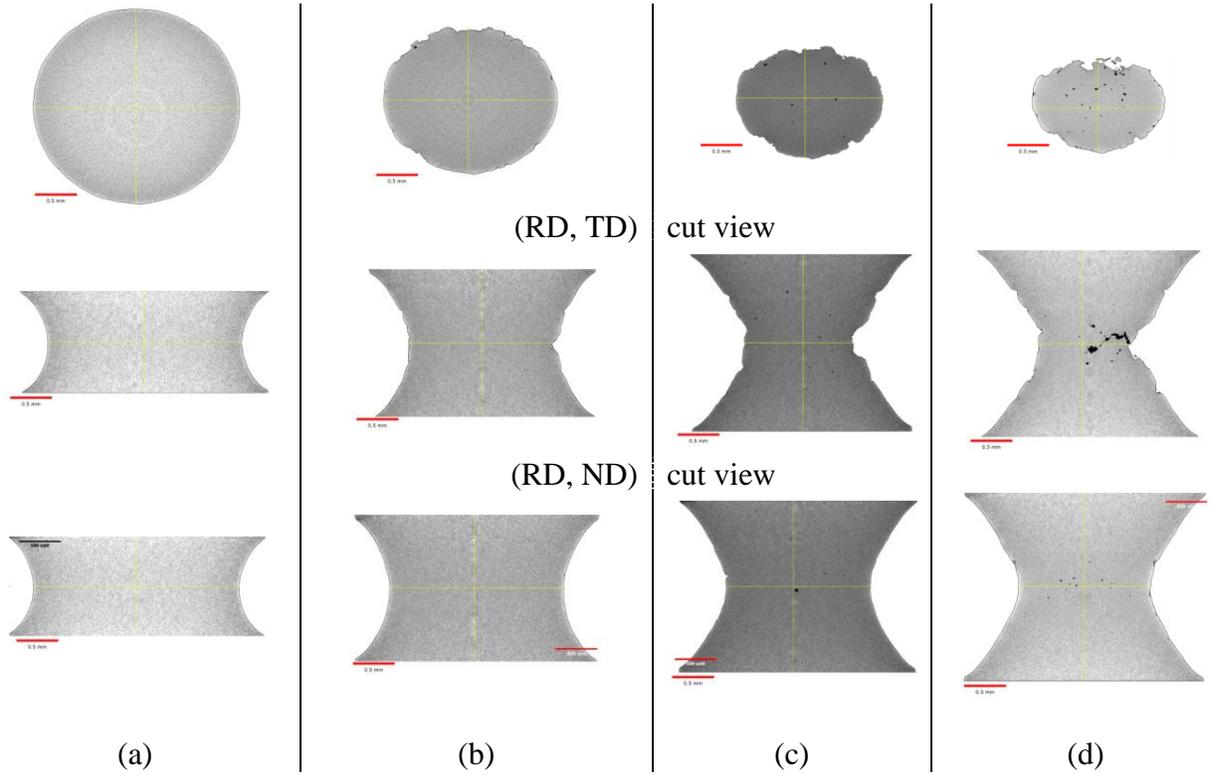

**Fig**. 14. X-ray micro-tomography in-situ scans of the HCP-titanium notched specimen subject to uniaxial tension along the rolling direction (RD). The (TD, ND) view, (RD-TD) view, and (RD, ND) views correspond to a notch displacement of : (a) 0.24 mm (b) 0.73 mm (c) 1.02 mm, (d) 1.2 mm. The rolling, transverse, and normal directions are denoted as RD, TD, ND.

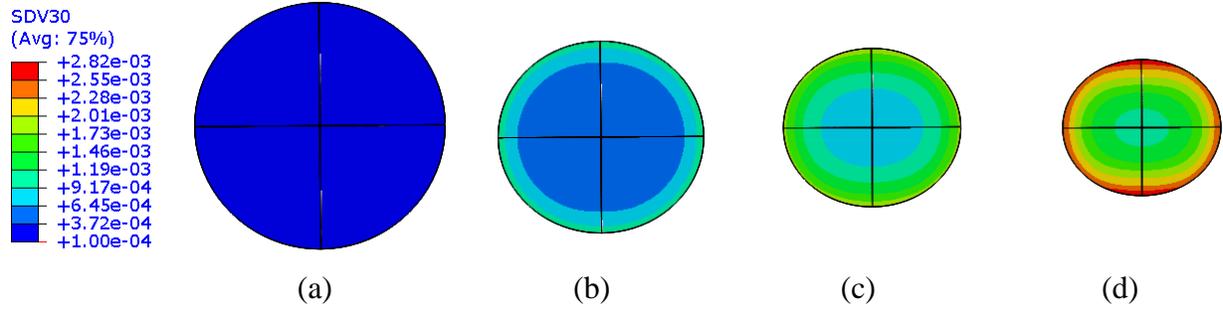

**Fig**. 15. F.E. predictions of the void volume fraction isocontours in the (TD, ND) cross-section of an axisymmetric notched specimen of HCP-titanium subjected to uniaxial tension along RD according to the Stewart and Cazacu (2011) model corresponding to an axial displacement of: (a) 0.24 mm (b) 0.73 mm (c) 1.02 mm (d) 1.2 mm, respectively. Initial void volume fraction $f_0=10^{-4}$. The rolling, transverse, and normal directions are denoted as RD, TD, ND.

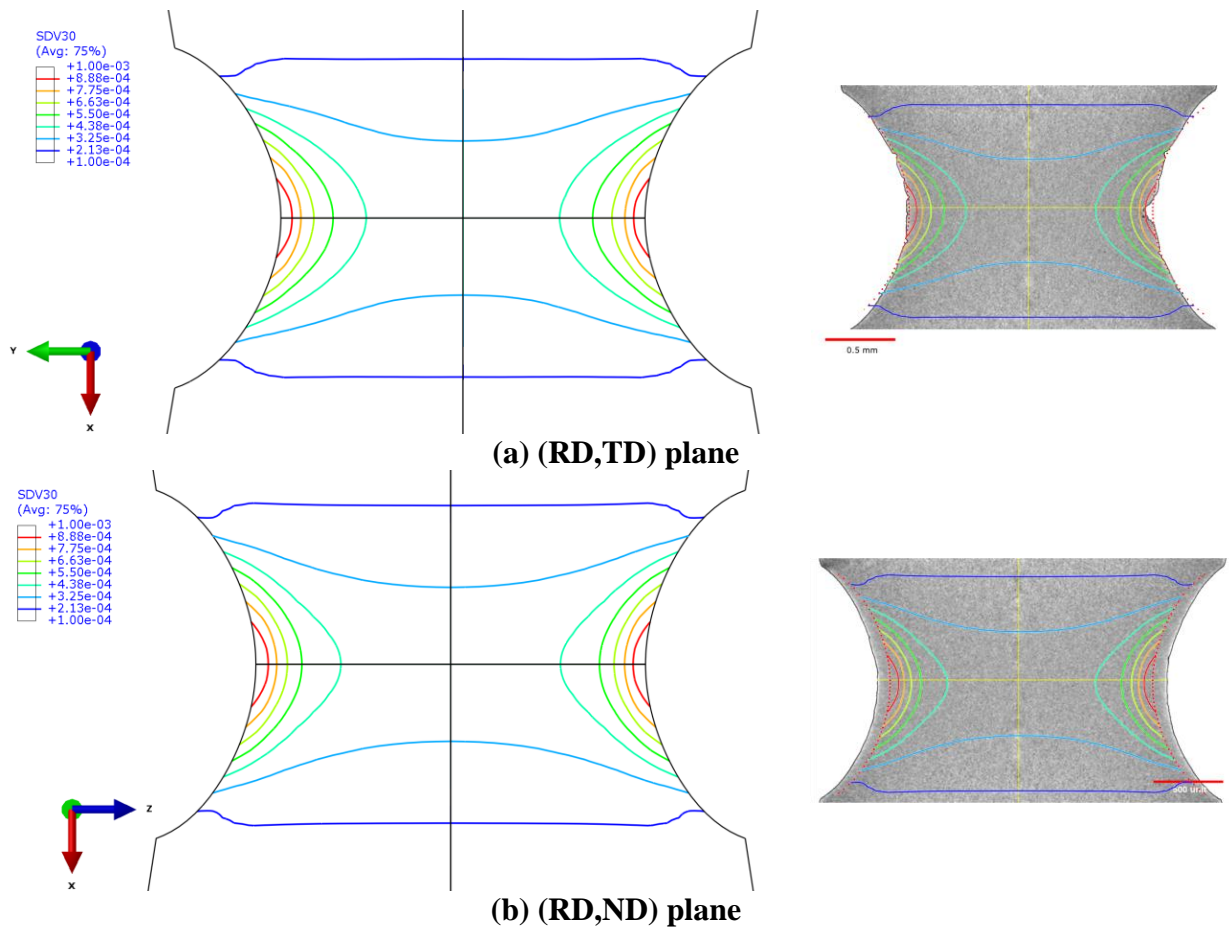

**Fig**.16. Comparison between the F.E. cross-sections and isocontours of void volume fraction of a notched axisymmetric specimen of HCP-titanium subjected to uniaxial tension along RD, according to the Stewart and Cazacu (2011) model and XCMT data for an axial notch displacement of 0.73 mm: (a) cross-section of the deformed specimen; (b) (RD-TD) section of the deformed specimen; (c) (RD-ND) section of the deformed specimen. The rolling,transverse, and normal directions are denoted as RD, TD, ND.

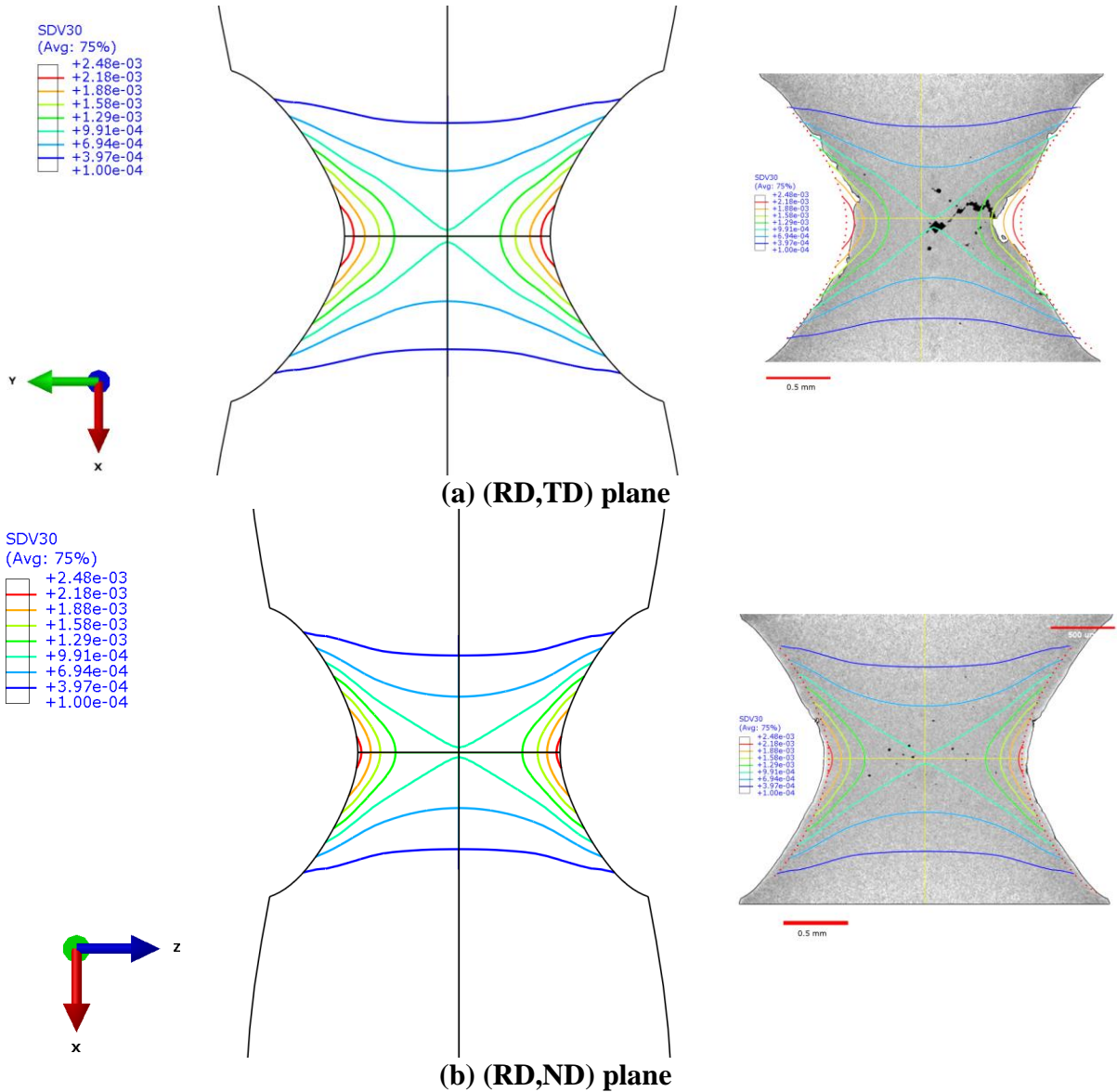

**Fig**.17. Comparison between the F.E. cross-sections and isocontours of void volume fraction of a notched axisymmetric specimen of HCP-titanium subjected to uniaxial tension along RD, according to the Stewart and Cazacu (2011) model and XCMT data for an axial notch displacement of 1.2 mm: (a) cross-section of the deformed specimen; (b) (RD-TD) section of the deformed specimen; (c) (RD-ND) section of the deformed specimen. The rolling,transverse, and normal directions are denoted as RD, TD, ND.